%% file: ms.tex
\documentclass[manuscript]{aastex}

\slugcomment{.}

\shorttitle{}
\shortauthors{}

\begin{document}


\title{350 Micron Dust Emission from High Redshift Quasars}


\author{Alexandre Beelen}
\affil{Radioastronomisches Institut der Universit\"at Bonn, 
  Auf dem H\"ugel 71, D-53121 Bonn, Germany}
\affil{Max-Planck-Institut f\"ur
  Radioastronomie, Auf dem H\"ugel 69, D-53121 Bonn, Germany}
\affil{Institut d'Astrophysique Spatiale, B\^at. 121,
  Universit\'e de Paris XI, F-91405 Orsay, France}
\email{abeelen@astro.uni-bonn.de}
\and
\author{Pierre Cox}
\affil{Institut de Radioastronomie Millimetrique, 
  St. Martin d'Heres, F-38406, France}
\email{cox@iram.fr}
\and
\author{Dominic J. Benford}
\affil{NASA / Goddard Space Flight Center, Code 665,
  Greenbelt, MD 20771}
\email{Dominic.Benford@nasa.gov}
\and
\author{C. Darren Dowell}
\affil{Jet Propulsion Laboratory, Mail Code 169-506, Pasadena, CA 91109}
\email{cdd@submm.caltech.edu}
\author{Attila Kov\'{a}cs}
\affil{Caltech Institute of Technology, MS 320-47, Pasadena, CA 91125}
\email{attila@submm.caltech.edu}
\and
\author{Frank Bertoldi} 
\affil{Radioastronomisches Institut Universit\"at Bonn, 
  Auf dem H\"ugel 71, 53121 Bonn}
\email{bertoldi@astro.uni-bonn.de}
\and
\author{Alain Omont} 
\affil{Institut d'Astrophysique de Paris, CNRS, and Universit\'e de Paris VI, 
  98bis boulevard Arago, F-75014 Paris, France}
\email{omont@iap.fr}
\and
\author{Chris L. Carilli}
\affil{National Radio Astronomy Observatory, P.O. Box, 
  Socorro, NM~87801, USA}
\email{ccarilli@aoc.nrao.edu}

\begin{abstract}
  
  We report detections of six high-redshift ($1.8 \le z \le 6.4$), optically
  luminous, radio-quiet quasars at 350~\micron, using the {\sc SHARC II}
  bolometer camera at the Caltech Submillimeter Observatory.  Our
  observations double the number of high-redshift quasars for which
  350~\micron\ photometry is available. By combining the 350~\micron~
  measurements with observations at other submillimeter/millimeter
  wavelengths, for each source we have determined the temperature of the
  emitting dust (ranging from 40 to $60\ \mathrm{K}$) and the far-infrared
  luminosity (0.6 to $2.2 \times 10^{13} \ \mathrm{L_\odot}$). The combined
  mean spectral energy distribution (SED) of all high-redshift quasars with
  two or more rest frame far-infrared photometric measurements is best fit with a
  greybody with temperature of $47 \pm 3\ \mathrm{K}$ and a dust emissivity
  power-law spectral index of $\beta = 1.6 \pm 0.1$.  This warm dust
  component is a good tracer of the starburst activity of the quasar host
  galaxy. The ratio of the far-infrared to radio luminosities of infrared
  luminous, radio-quiet high-redshift quasars is consistent with that found
  for local star-forming galaxies.

\end{abstract}



\keywords{infrared: QSOs, galaxies --- galaxies: starburst, evolution,
  active quasars: individual (\objectname{KUV~08086+4037},
  \objectname{APM~08279+5255}, \objectname{HS~1002+4400},
  \objectname{SDSS~J1148+5251}, \objectname{J1409+5628},
  \objectname{PSS~2322+1944}) --- infrared: ISM: continuum}


\section{Introduction}
\label{sec:introduction}

At high redshifts, a large fraction of the star formation occurs in very
luminous ($> 10^{12}\ \mathrm{L_\odot}$) galaxies \citep{Greve2005}.  Most
of the radiation of the newly formed stars in these systems is absorbed by
the dust and re-emitted at infrared (IR) wavelengths. In their rest frame,
the spectral energy distribution (SED) of dusty galaxies typically peaks at
60-80~\micron~ and can be approximated with a modified blackbody spectrum
with dust temperatures between 30 and 80~K \citep{Dunne2000}. In host
galaxies of quasi-stellar objects (QSOs), some dust can be directly heated
by the central Active Galactic Nucleus (AGN) to higher temperatures; its
emission then dominates the near- to mid-IR SED.  For sources at high
redshift ($z>1$), the peak of the emission is shifted to submillimeter
wavelengths, enabling useful multi-color photometric observations with
ground-based telescopes in the submillimeter (submm) and millimeter (mm)
atmospheric windows.  Tracing the peak of their SED is crucial for an
accurate determination of the IR luminosity and star formation rate in such
dusty, energetic systems.

In principle, a determination of the SED peak could be expected to be useful
for estimating the source redshift in cases where spectroscopic redshifts
are not available. While this method has been shown to work in limited Monte
Carlo simulations \citep{Wiklind2003, Hughes2002}, there is the distinct
possibility that it may be impractical, as the redshift and dust temperature
are largely degenerate \citep{Blain2003}.  Sampling the SED of high-redshift
IR-luminous galaxies through the several submm and mm atmospheric windows
(usually 350, 450, 850\micron, 1.2, 2.0, 3.0~mm) has been achieved for only
a few high-$z$ sources \citep[][ and references therein]{Benford1999,
  Priddey2001}.

Here we present observations at 350$\,\mu$m of six quasars in the
redshift range $1.8\lesssim z\lesssim 6.4$. This doubles the number of
high-$z$ quasars for which such measurements are available.  The
optically luminous and radio-quiet quasars were selected from the
1.2~mm continuum surveys performed with the Max-Planck
Millimeter Bolometer (\emph{MAMBO}) array by Omont et al. (2001, 2003)
and Bertoldi et al. (2003a) - see Table~\ref{tab:observation}.  The
selected wavelength of 350$\,\mu$m roughly corresponds to the peak in
the SEDs of highly redshifted emission from dust with temperatures of
40 to 60$\,$K.  We adopt the concordance $\Lambda$-cosmology with
$H_0=71\ \mathrm{km\ s^{-1}\ Mpc^{-1}}$, $\Omega_\Lambda=0.73$ and
$\Omega_m=0.27$ \citep{Spergel2003}.

\section{Observations}
\label{sec:observations}

The measurements were made on January 6 and 7, 2004 at the 10.4~m Leighton
telescope of the Caltech Submillimeter Observatory (CSO) on the summit of
Mauna Kea, Hawaii, during excellent weather conditions, with stable zenith
atmospheric opacities of 1.0 at 350~\micron.  We used the CSO bolometer
camera, {\sc SHARC II}, described by \citet{Dowell2003}, moounted at the
reimaged Cassegrain focus.  It consists of a $12\times 32$ array of doped
silicon 'pop-up' bolometers operating at 350~\micron.  Under good weather
conditions, the point-source sensitivity at 350~\micron~ is $\sim 1\
\mathrm{Jy/\sqrt{Hz}}$ or better. The detectors of {\sc SHARC II} cover the
focal plane with 90\% filling factor and are separated by $0.70 \, \lambda/D
= 4.86\arcsec$.

Pointing was checked regularly on strong sources including planets and
secondary calibrators, and the focus was checked at the beginning of each
night. The pointing was found to be stable with a typical accuracy of $
\lesssim 2^{\prime\prime}$ ($1\sigma$ r.m.s). We used the Dish Surface
Optimization System (DSOS) which actively corrects the 10.4-meter primary
surface for static imperfections and deformations due to gravitational
forces as the dish moves in elevation \citep{Leong2005}. The DSOS improves
the beam shape and provides an elevation-independent telescope efficiency
$\sim$~10\% better than the passive telescope (which has an aperture
efficiency of 33\% at 350~\micron) at intermediate elevation and 50\% better
at high elevation.  The beam of the CSO at 350~\micron~ is approximated by a
circular Gaussian with a FWHM of $8\farcs5$.  Uranus served as a primary
flux calibrator and the asteroids Pallas and Ceres as secondary calibrators.
At the time of the observations, the brightness temperature of Uranus was
64~K and its diameter $3\farcs4$, corresponding to a 350~\micron~ flux
density of 224~Jy. The secondary calibrators Pallas, Ceres and IRC10216 had
350~\micron~ flux densities of 7.0, 40 and 24~Jy, respectively, as derived
by various observation made with {\sc SHARC II} using Mars, Uranus and
Neptune as primary calibrators, with uncertainties of 15\% based on
repeatability of observations.  The absolute calibration was found to be
accurate to within 20\% at 95\% confidence.

{\sc SHARC II} was also used at 450~\micron~ to observe APM~08279+5255 on
January 13, 16 and 18, 2004 under good weather conditions
($\tau_\mathrm{225\ GHz} \sim 0.09$).  For these observations, we used as
the secondary calibrators Ceres and IRC+10216 which had flux densities at
450~\micron~ of 28 and 13~Jy, respectively.

The observations were performed by scanning the {\sc SHARC II} array in
azimuth and elevation using Lissajous and box-scan patterns with amplitudes
of $15\farcs0 \times 14\farcs14$ and $25\farcs0 \times 35\farcs22$,
respectively, in order to reduce systematic errors. The {\sc SHARC II} array
is rectangular, of $2\farcm61 \times 0\farcm96$ in extent, with the long
axis oriented in azimuth. The sources were also observed at different
elevations before and after transit.  The final maps have typical sizes of
$\approx 2 \times 1\ \mathrm{arcmin^2}$.  At 350~\micron, the total
integration time per map is between 150 and 430 minutes, with the exception
of APM 08276+5255, which was only observed for only 20 minutes.  The
corresponding rms map noise values are in the range from 5 to 15~mJy
(Table~\ref{tab:observation}).

The data were reduced using the 1.42 version of the software package
\emph{CRUSH} (Comprehensive Reduction Utility for {\sc SHARC II}; Kov\`{a}cs
2006, in preparation). \emph{CRUSH} is based on an algorithm that solves a
series of models which try to reproduce the observations through an
iterative process, taking into account instrumental and atmospheric effects
while using statistical estimators on the data. For the current data, we
used the option \emph{deep} which is appropriate for sources with typical
flux densities smaller than 1~Jy/beam.  To derive the flux densities of the
high-$z$ quasars at 350 and 450~\micron, we fitted circular Gaussians
profiles and a uniform background level, leaving the position, the full
width half maximum and the intensities as free parameters. The resulting
noise maps were used to derive the statistical uncertainties on the fitted
parameters which served to estimate the uncertainty on the flux densities.
These uncertainties do not take into account the calibration errors.

\section{Results}
\label{sec:results}

\begin{figure}
\plottwo{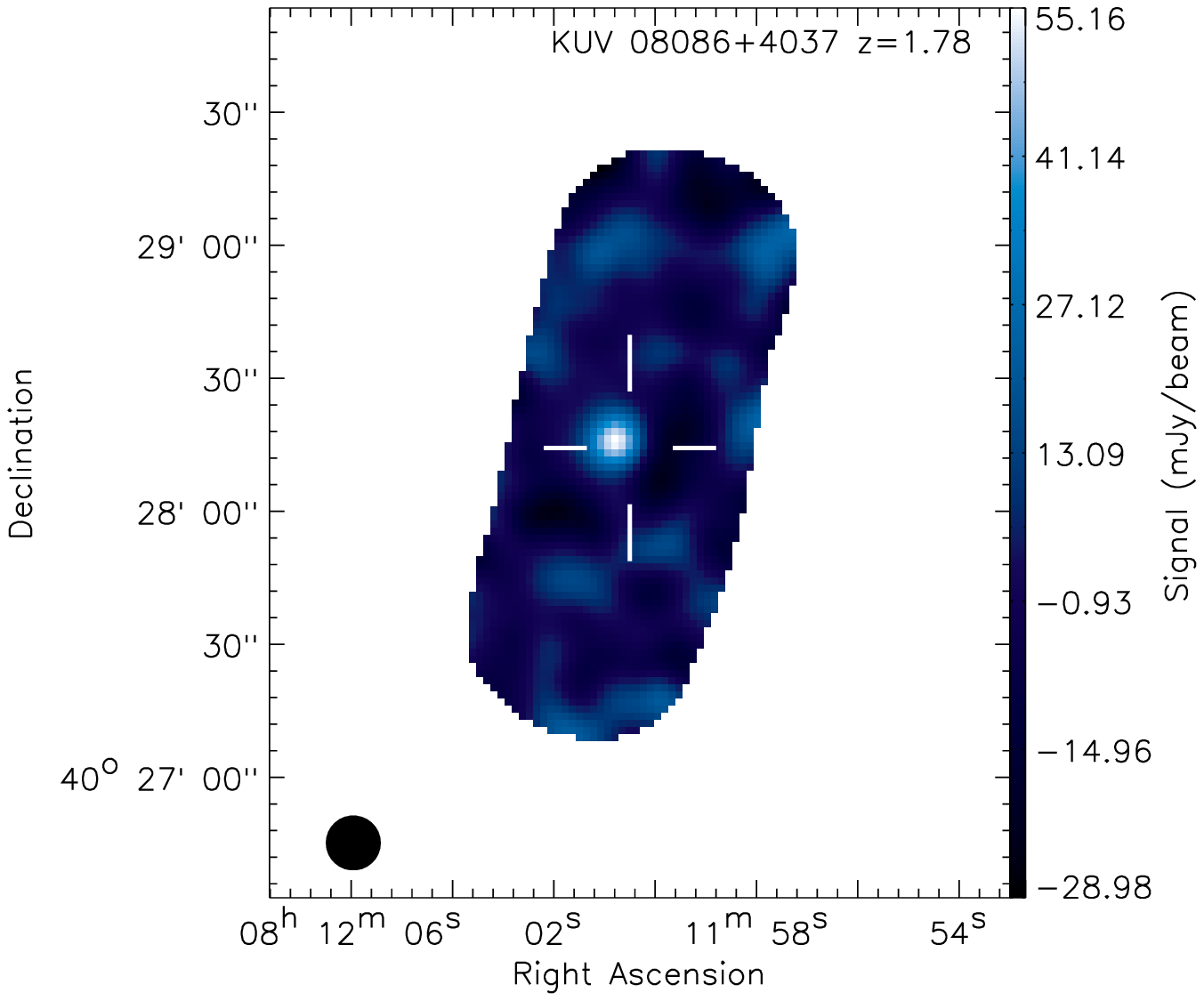}{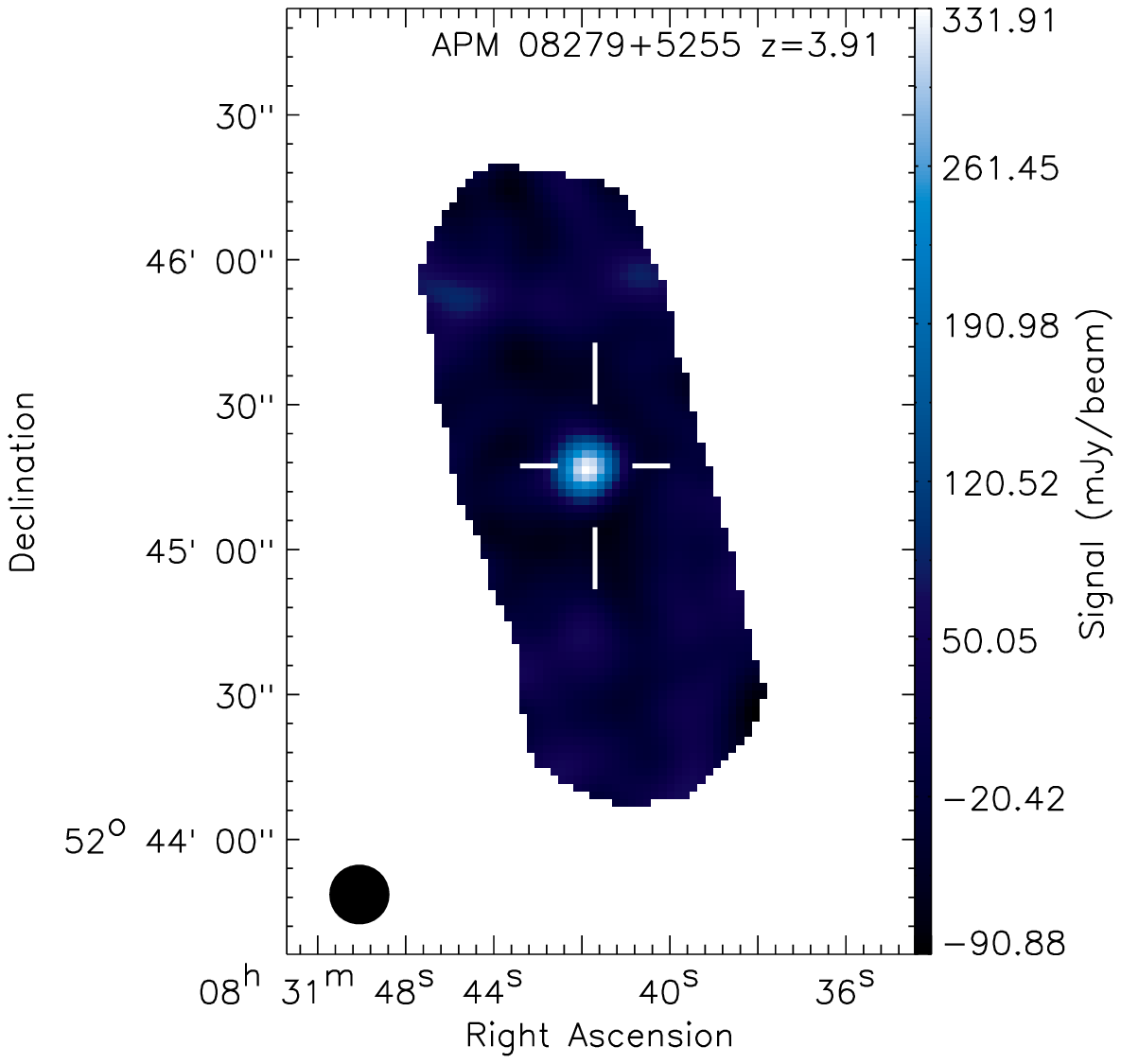} \\
\plottwo{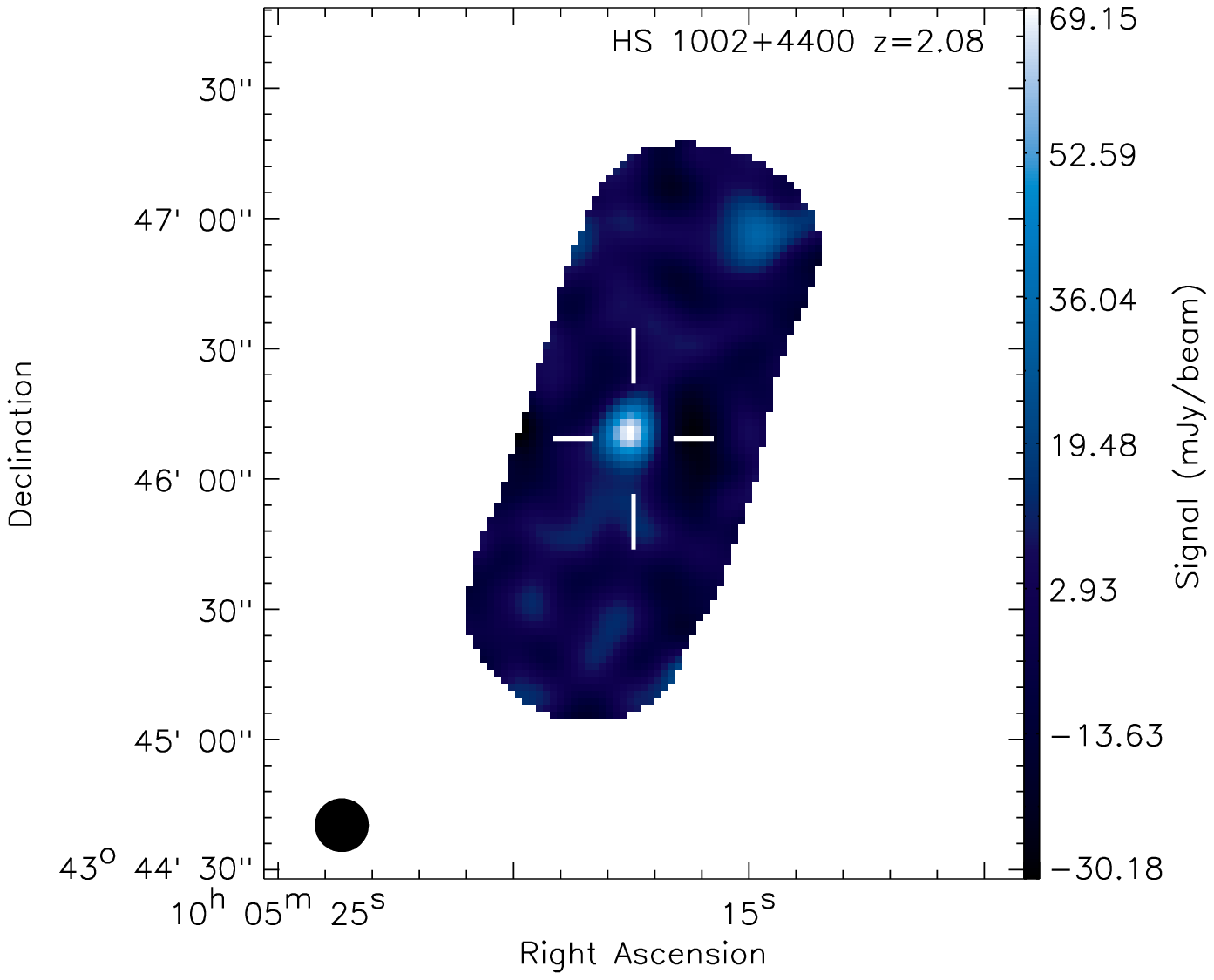}{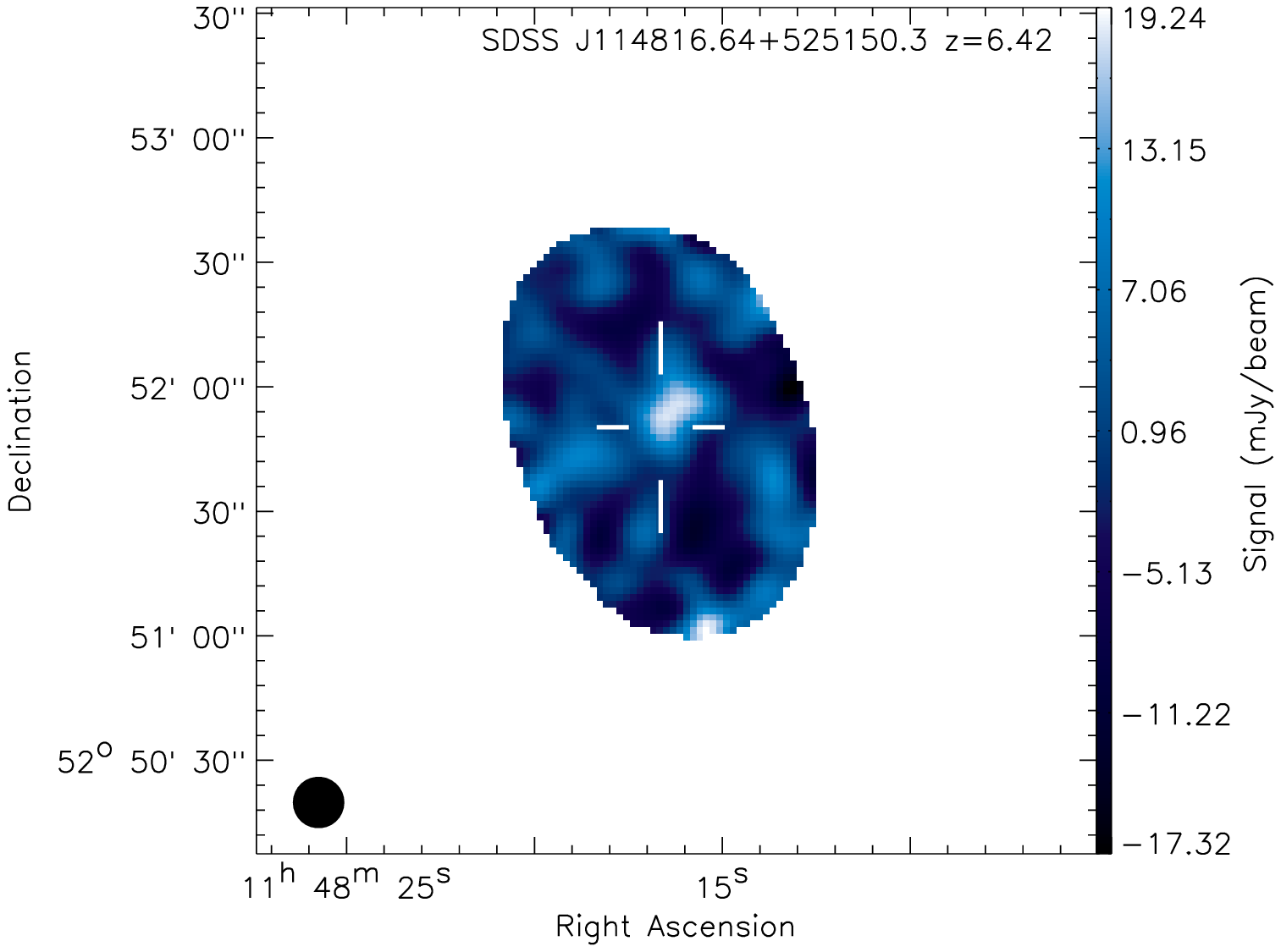} \\
\plottwo{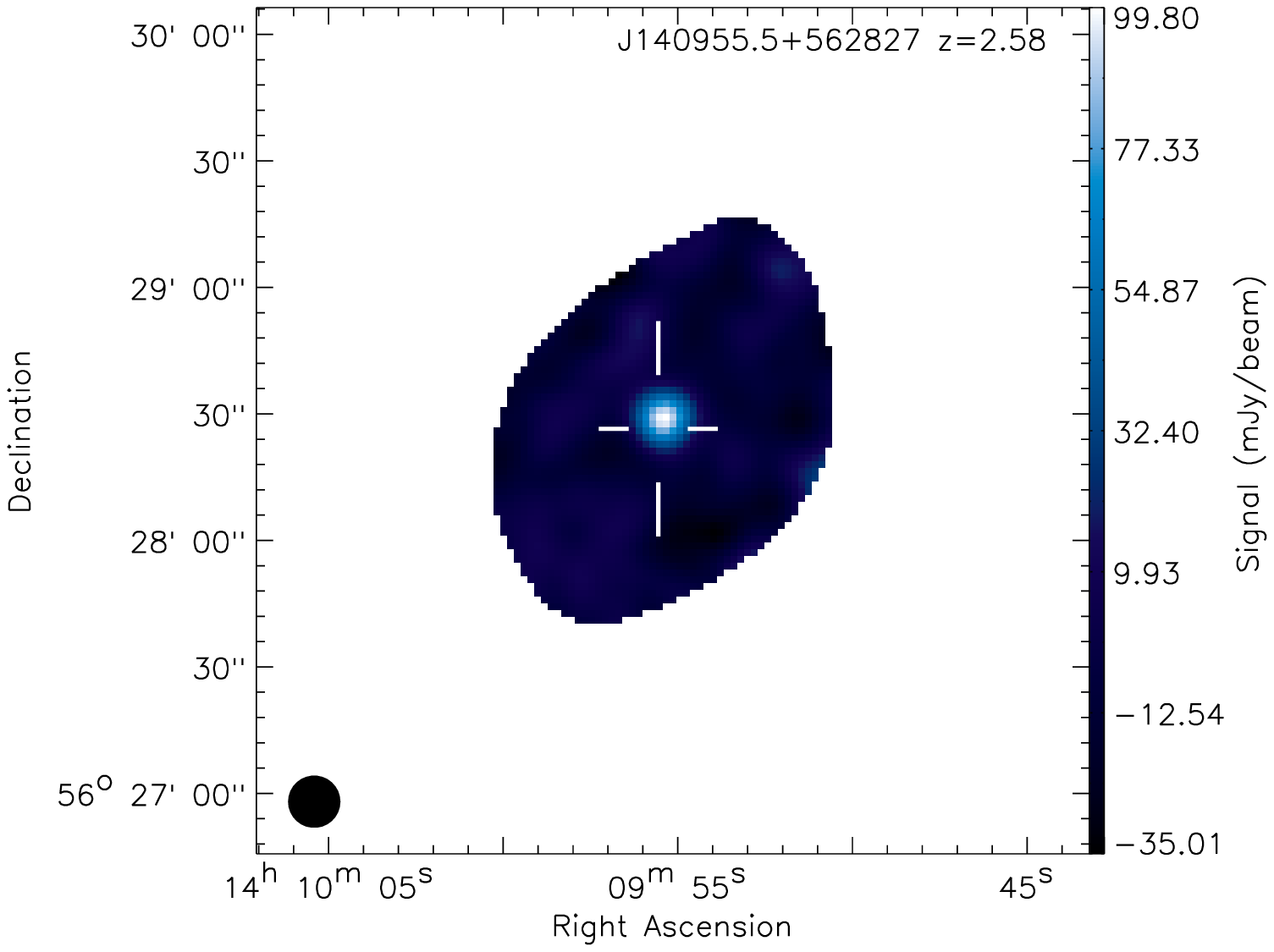}{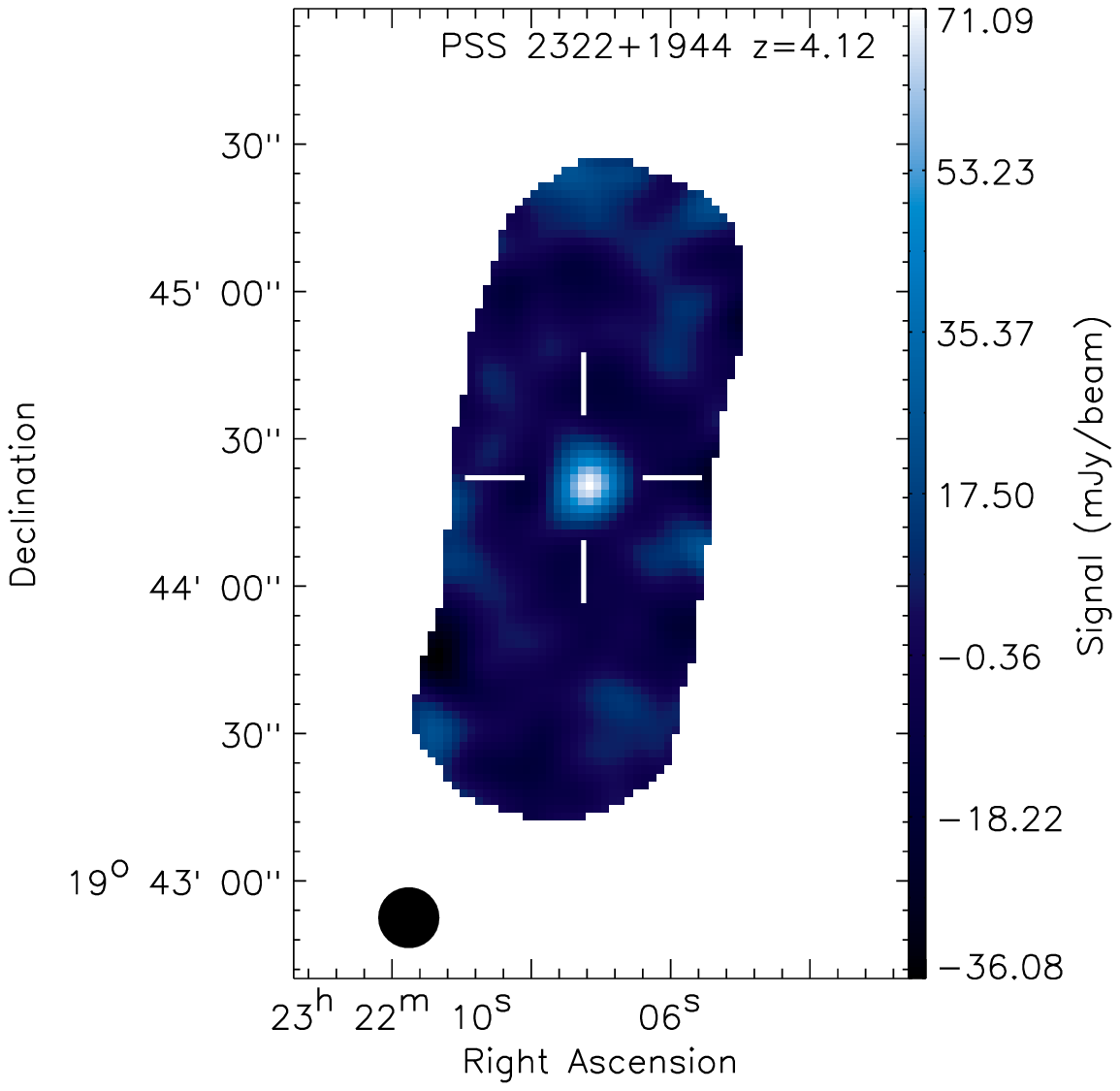} 
\caption{{\sc SHARC II} maps at 350~$\mu$m of 
  the six optically luminous, radio-quiet high-$z$ quasars studied in
  this paper. The names and redshifts are given in the upper left
  corner of each panel. The maps have been cut to half of the 
  maximum exposure, and smoothed with a 9\arcsec\ FWHM Gaussian.The
  resulting beam size is shown in the lower left corner of each plot.
  The crosses indicate the optical positions of the quasars as listed in
  Table~\ref{tab:observation}.
  \label{fig:maps}}
\end{figure}

All six quasars were detected at significance levels of $>\sim3.5\,\sigma$
(Table~\ref{tab:observation}). The signal maps of the 350~\micron~ continuum
emission are shown in Fig.~\ref{fig:maps}.  In the following, we comment on
the individual sources.

{\bf KUV~08086+4037} This $z=1.78$ broad emission line quasar was discovered
by \citet{Darling1996} and detected at 1.2~mm by \citet{Omont2003}. The mm
to 1.4~GHz radio continuum flux ratio is compatible with a star-forming
galaxy Petric et al. (2006, in preparation).  We detect KUV~08086+4037 at 350~\micron~ with a
flux density of $69 \pm 11\ \mathrm{mJy}$. The peak emission of
KUV~08086+4037 is shifted by 3 to $\sim$4\arcsec\ from the optical or radio
position, corresponding to $\sim2\sigma$ of the pointing accuracy of the CSO.
However, it is unlikely that the detected emission does not come from the
QSO, and the offset is probably due to a pointing model problem.

{\bf APM~08279+5255} This is a strongly lensed $z=3.91$ broad absorption
line quasar with a bolometric luminosity of $\sim 10^{14}\ \mathrm{L_\odot}$
after correction for magnification by a factor 7 \citep{Downes1999}.  It is
a strong IR/submm source which was even detected by IRAS \citep{Irwin1998}.
Its mid-IR spectrum was observed with the {\em Spitzer Space Telescope},
confirming a strong continuum likely due to the dust heated by the AGN
\citep{Soifer2004}.  The massive reservoir of dust (a few $\sim 10^8 \,
\mathrm{M_\odot}$) and warm molecular gas ($3 \times 10^9\
\mathrm{M_\odot}$) associated with this quasar is distributed in a nuclear
disk of radius 100-200~pc around the Active Galactic Nucleus (AGN), and
traces a dynamical mass of $1.5 \times 10^{10} \, \mathrm{M_\odot}$
\citep{Lewis2002}.  With a 350~\micron~ flux density of $386 \pm 32\
\mathrm{mJy}$, APM~8279+5255 is the strongest high-$z$ source ever detected
at 350~\micron.  The flux density at 450~\micron~ measured with {\sc SHARC
  II} is $342 \pm 26\ \mathrm{mJy}$, and were found to be stable over the
three different observations periods, $319\pm32\ \mathrm{mJy}$ on January
the 13rd, $279\pm111\ \mathrm{mJy}$ on the 16th and $418\pm52\ \mathrm{mJy}$
on the 18th. This flux density is higher than the $211 \pm 47\ (68)\
\mathrm{mJy}$ reported by \cite{Lewis1998} and the $285 \pm 11\ (40)\
\mathrm{mJy}$ found by \cite{Barvainis2002}, but is still compatible with
them within $1\sigma$, when taking into account the absolute calibration
uncertainties of {\sc SCUBA}, as quoted in parenthesis and estimated to be
of the order of 10\%.
  
{\bf HS~1002+4400} is a quasar at $z=2.08$ which was first discovered in the
Hamburg survey \cite[]{Hagen1999}. It displays broad emission lines with no
peculiar features, and it was detected at 1.2~mm by \cite{Omont2003} and in
the 1.4~GHz radio continuum by \citet{Petric2005}. The radio to mm
continuum flux ratio is consistent with that of a starburst.  HS~1002+4400
is detected at 350~\micron~ with a flux density of $77 \pm 14\
\mathrm{mJy}$.
  
{\bf SDSS~J114816.64+5251}, hereafter refered to as J1148+5251, is the most
distant quasar ($z=6.42$) known to date \citep{Fan2003}. An optically very
luminous quasar, powered by a supermassive black hole, it also shows strong
far-IR emission with an estimated luminosity of $1.2 \times 10^{13}\
\mathrm{L_\odot}$ \citep{Bertoldi2003a}, and an implied rate of star
formation of $\approx 3000\ \mathrm{M_\odot\ yr^{-1}}$.  A massive reservoir
of dense molecular gas ($\approx 2 \times 10^{10} \ \mathrm{M_\odot}$) is
implied by the observed CO emission \citep{Walter2003, Bertoldi2003b} and
the recent detection of the [CII] emission line \citep{Maiolino2005}.  The
1.4~GHz radio continuum flux follows the radio-FIR correlation for
star-forming galaxies \citep{Carilli2004}.  J1148+5251 as been detected both
at 850 and 450\micron by \citet{Robson2004}, with flux densities of
$7.8\pm0.7\ \mathrm{mJy}$ and $24.7\pm7.4\ \mathrm{mJy}$ respectively.
\citet{Charmandaris2004} reported the detection of J1148+5251 with the {\em
  Spitzer Space Telescope} at 16 and 22~\micron, revealing a hot dust
component that could be heated by the AGN; however, it should be noted that
these measurement are also compatible with the emission of the AGN
extrapolated from the optical.  We detected J1148+5251 at 350~\micron~ with
a flux density of $21 \pm 6 \ \mathrm{mJy}$ at the optical position of the
QSO.  This is the lowest flux density ever detected at 350~\micron. A map
with the best fit source subtracted present only noise, with any residual
sources $<14\ \mathrm{mJy}$, meaning that the apparent source extension seen
in figure~\ref{fig:maps} is due to noise.

{\bf J140955.5+562827}, hereafter refered to as J1409+5628, at $z=2.56$ is
an optically very bright, radio-quiet quasar. It is by far the strongest mm
source in the \citet{Omont2003} MAMBO survey of $z \approx 2$ quasars. It
has a massive reservoir of warm and dense molecular gas which is detected in
CO \citep{Beelen2004} and HCN \citep{Carilli2005}, with an estimated mass of
$6 \times 10^{10} \ \mathrm{M_\odot}$.  Its far-IR luminosity of $\sim 4
\times 10^{13}\mathrm{L_\odot}$ implies a star formation rate of several
$1000 \ \mathrm{M_\odot \ yr^{-1}}$ \citep{Beelen2004}.  The radio flux
density is consistent with the radio-FIR correlation for star-forming
galaxies Petric et al. (2006, in preparation).  High resolution VLBA
observations resolve out the radio emission, implying an intrinsic
brightness temperature of $\sim 10^5 \ \mathrm{K}$ at 8~GHz, which is
typical for nuclear starbursts \citep{Beelen2004}.  The 350~\micron~ flux
density of $112 \pm 12 \ \mathrm{mJy}$ places J1409+5628 among the strongest
submm high-$z$ sources.

{\bf PSS~2322+1944} is an optically luminous, lensed (magnification $a=3.5$)
quasar at $z=4.12$. It was detected in mm dust and radio continuum, and in
various CO emission lines \citep{Omont2001, Cox2002, Carilli2002}. With a
far-IR luminosity of $9 \times 10^{12} \, L_\odot$ (corrected for lensing),
it harbours a massive reservoir of molecular gas detected in CO, where star
formation takes place with a rate of $\approx 1000 \ \mathrm{M_\odot\
  yr^{-1}}$.  The CO line emission is resolved into an Einstein ring, which
can be modeled as a disk of dense and warm molecular gas surrounding the QSO
with a radius of $\sim 2 \, \mathrm{kpc}$, tracing a dynamical mass of a few
$10^{10} \ \mathrm{M_\odot}$ \citep{Carilli2003}.  PSS~2322+1944 is clearly
detected at 350~\micron~ with a flux density of $79 \pm 11 \ \mathrm{mJy}$.

\section{Discussion}
\label{sec:discussion}

The thermal emission from dust at a temperature $T_\mathrm{dust}$, is
proportional to $B_\nu(T_\mathrm{dust})[1-e^{-\tau_d}]$, where
$B_\nu(T_\mathrm{dust})$ is the Planck function and $\tau_d(\lambda) =
\kappa(\lambda)\int \rho\ ds$ is the dust optical depth, with $\kappa$ being
the mass absorption coefficient at rest wavelength $\lambda$ and $\rho$ the
total mass density. At far-IR wavelengths $\lambda> 40$~\micron, the
emission is optically thin, $\tau_d \ll 1$, so that the above expression
applied to a given dust mass at redshift $z$ leads to an emergent flux
density
\begin{eqnarray}
\label{eq:mbb}
 S_{\nu_\mathrm{o}} &=& \frac{(1+z)}{{D_\mathrm{L}}^2}\ M_\mathrm{dust}\  
                        \kappa(\nu_\mathrm{r})\
 B_{\nu_\mathrm{r}}(T_\mathrm{dust})\\ 
 &\propto& {\nu_\mathrm{r}^{3+\beta}}\ 
 \frac{1}{\exp(h \nu_\mathrm{r}/k T_\mathrm{dust}) - 1 },
\end{eqnarray}
where the mass absorption coefficient, $\kappa(\nu) = \kappa_0
(\nu/\nu_0)^\beta$, is approximated by a power law.

Dusty galaxies typically show multi-temperature components in their IR SEDs.
At rest frame near- and mid-IR wavelengths, the emission from these quasars
arises from a ``hot'' (several 100~K) but not very massive dust component,
that is likely heated directly by the AGN.  The far-IR emission, on the
other hand, arises from starburst regions where most of the energy emerges
from dust with temperatures in the range 30 to 80~K, but by mass, most of
the dust could be in a 10-20~K component that is energetically overpowered
by the warmer component.

Through a fit of a single temperature greybody (3 free parameters:
luminosity or integrated intensity, $\beta$ and $T_\mathrm{dust}$), the dust
temperature and emissivity index can be determined simultaneously only if
{\em at least} 4 photometric data points are available (fit with more than
one degree of freedom). For most objects, only two photometric data points
($S_1$ and $S_2$ at $\nu_1$ and $\nu_2$) are available, resulting in a
degeneracy between the dust temperature and $\beta$
\citep{Priddey2001,Blain2003}:
\begin{equation}
  \label{eq:degeneracy}
  \alpha e^{h \nu_1/ k T_\mathrm{dust}} - e^{h \nu_2/k
  T_\mathrm{dust}} = \alpha - 1,
\end{equation}
where $\alpha = \left({\nu_2/\nu_1}\right)^{3+\beta}\ S_1/S_2$.

\subsection{Individuals objects}
\label{sec:individual}

We fit a modified black (grey) body spectrum (Eq.~\ref{eq:mbb}) to the
rest frame 40 to 800~\micron\ SEDs of the six high-$z$ quasars we
observed at 350~\micron.  To account for the uncertainties in the
absolute calibration of the submm and mm photometry,
we added 20\% to the uncertainties of all measurements.

\begin{figure}
\plottwo{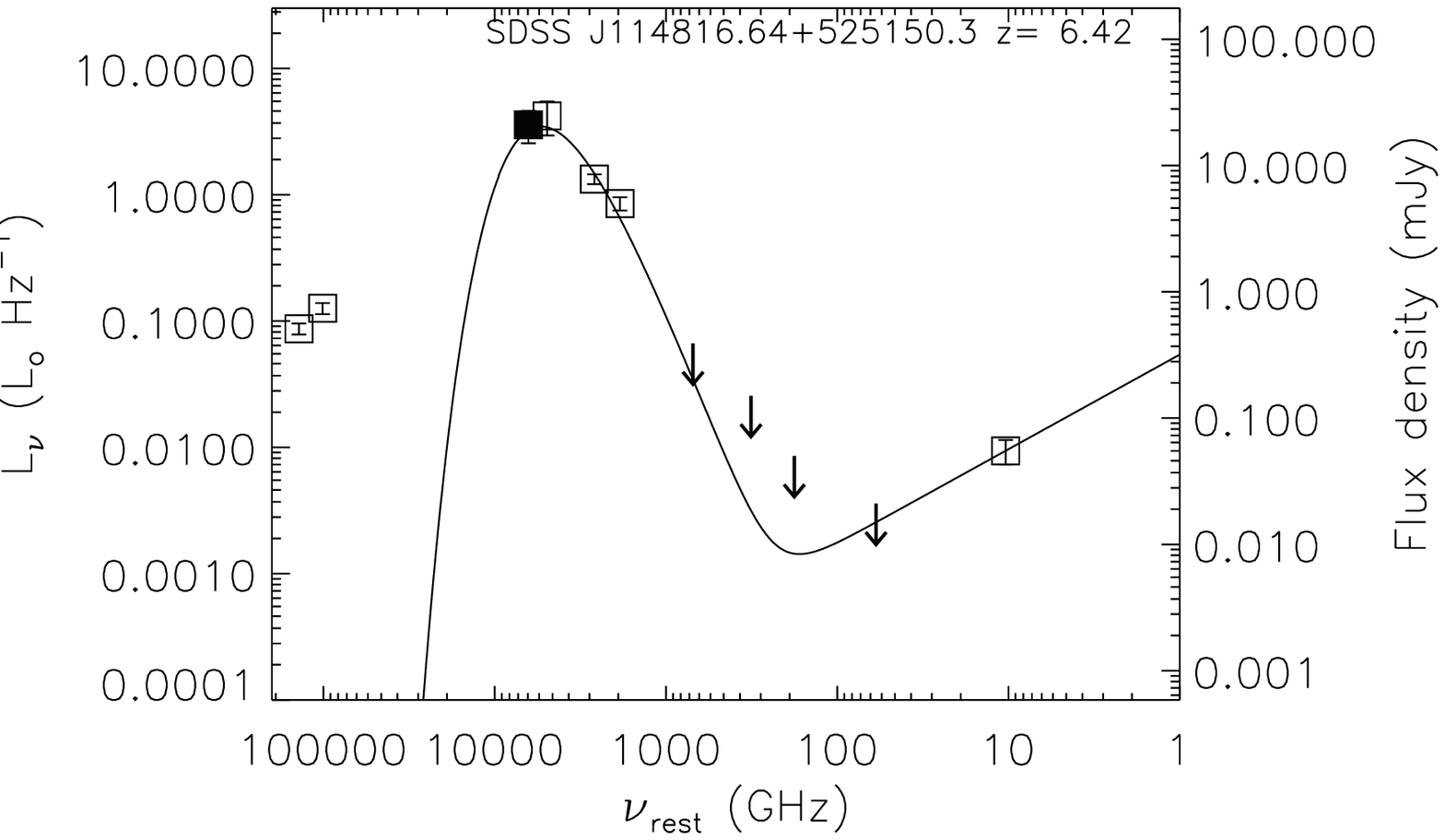}{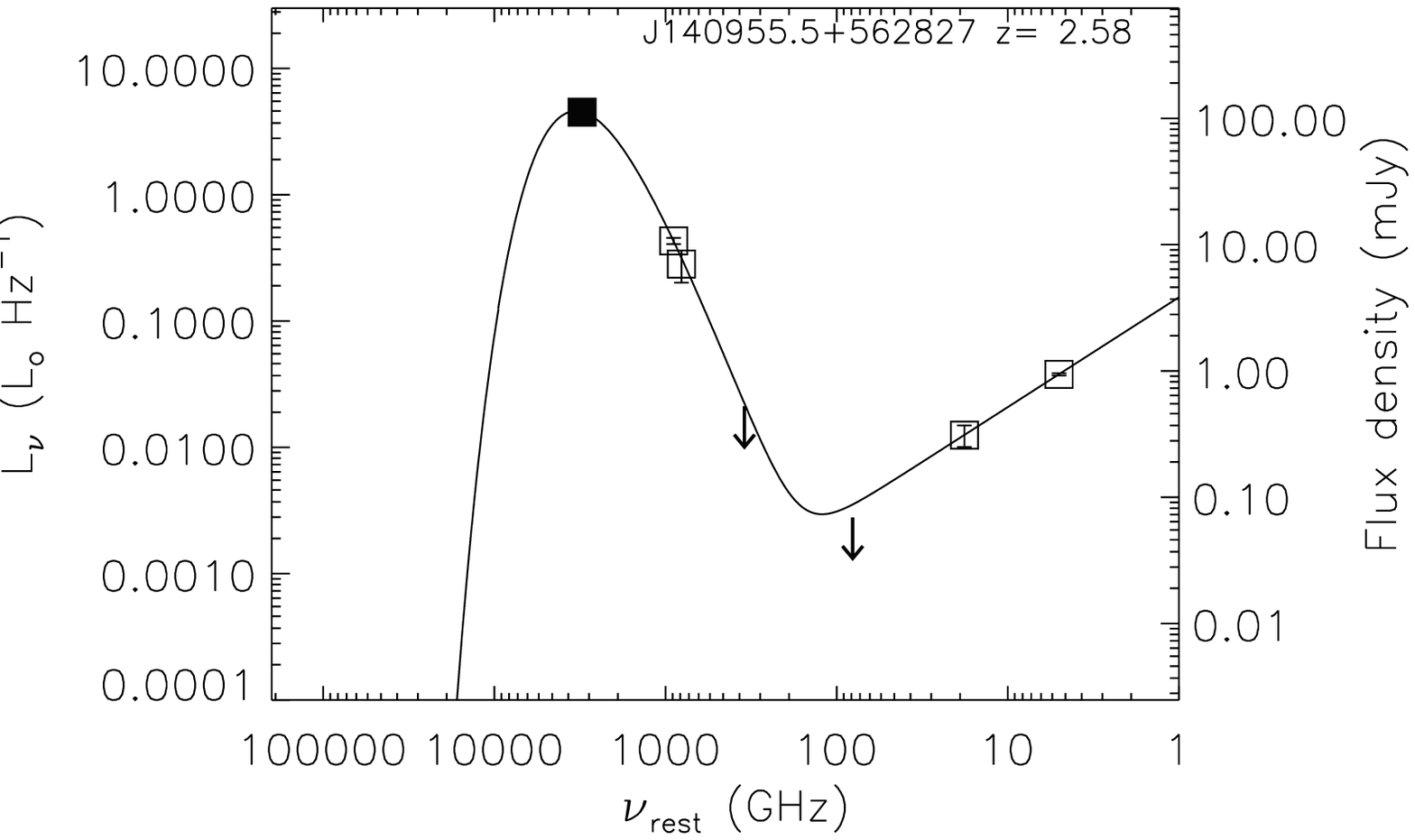} \\
\plottwo{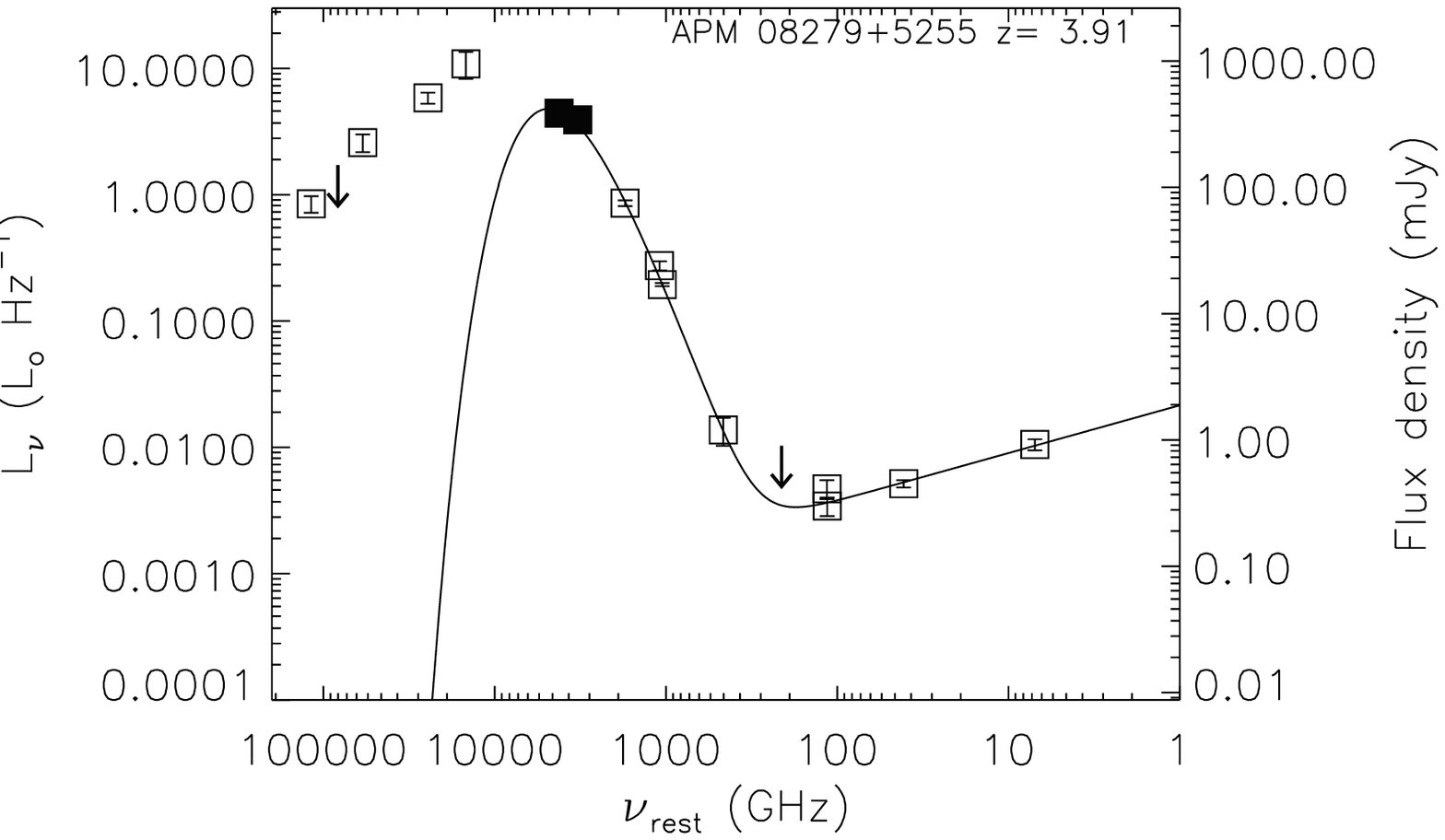}{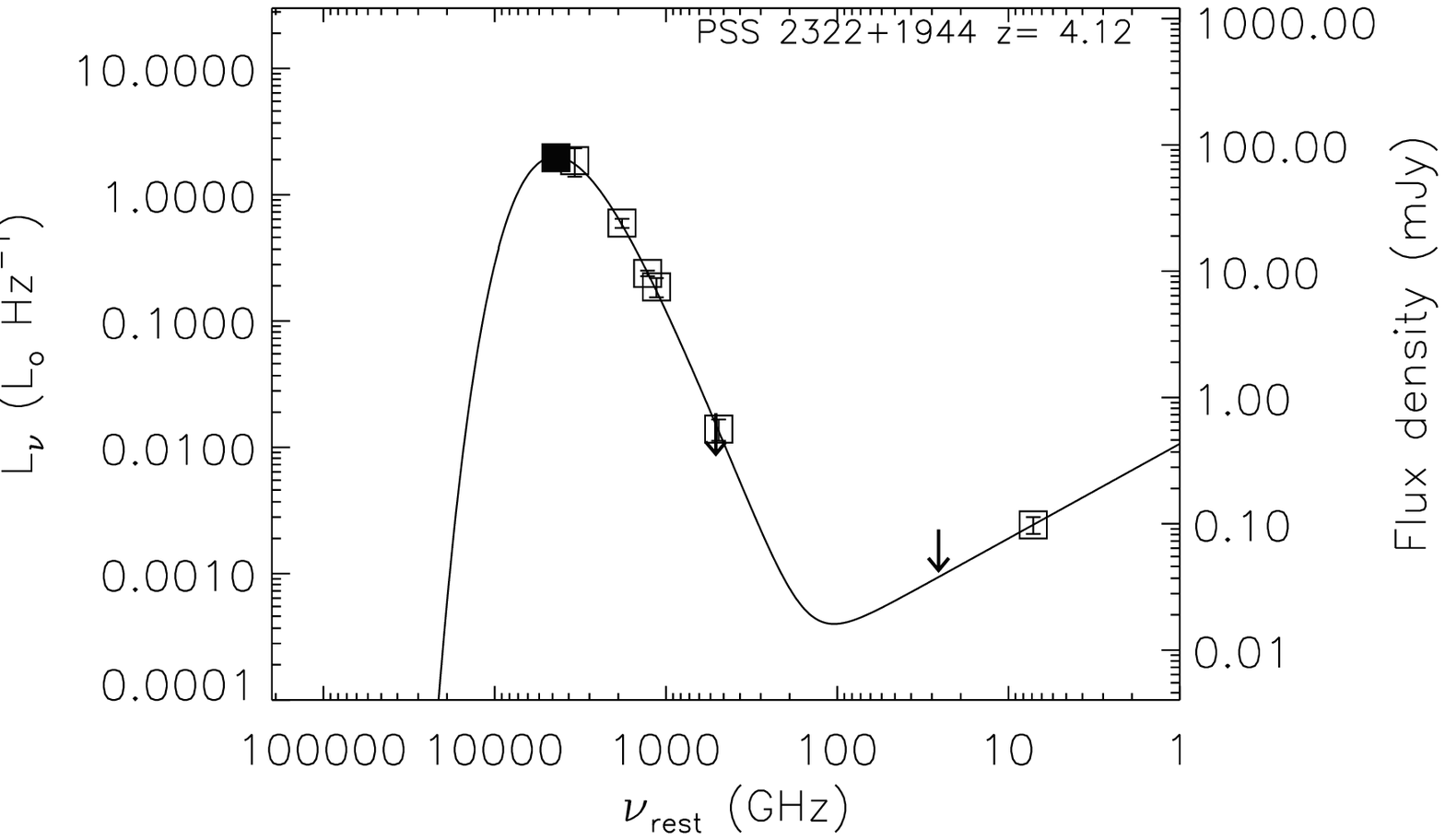}
\caption{Spectral energy distributions (SEDs) of J1148+5251, J1409+5628,
  APM~08279+5255 and PSS 2322+1944 in the rest frame of the sources.  The
  {\sc SHARC II} 350 and 450~\micron~ points are shown as filled squares.
  Other measurements (taken from the literature) are displayed with open
  squares with arrows indicating $3\sigma$ upper limits. 
  The solid line shows the best fit to the
  far-IR data using a modified black-body and to the radio using a
  power-law (see text and parameters in Table~\ref{tab:derived}). When
  needed the radio spectral index was set to 0.75. The references
  to the photometric measurements other than from {\sc SHARC II} are:
  J1148+5251 - \citet{Bertoldi2003a, Carilli2003, Carilli2004,
  Robson2004, Charmandaris2004}; J1409+5628 - \citet{Omont2003,
  Beelen2004, Petric2005}; PSS~2322+1944 \citet[][ and references
  therein]{Cox2002}; APM08279+5255 - \citet{Irwin1998, Lewis1998,
  Lewis2002, Downes1999}.
\label{fig:indiv_plots}}
\end{figure}

For those quasars where only two photometric data points are available, we
fixed $\beta = 1.6$ (see \S.~\ref{sec:mean}) and derived $T_\mathrm{dust}$
from Eq.~\ref{eq:degeneracy}. When more photometric measurements are
available, we performed a $\chi^2$-fit on $T_\mathrm{dust}$ only, or both
$T_\mathrm{dust}$ and $\beta$ when the degree of freedom in the fit was
greater than one, except in the case of J1148+5251 were the lack of
detection in the Rayleigh-Jeans domain prevent the fit of $\beta$. The
resulting fits are shown in Fig.~\ref{fig:indiv_plots} for four of the
quasars where three or more measurements are available.
Figure~\ref{fig:indiv_plots} also shows the available radio, near- and
mid-IR measurements, as well as the radio continuum emission expected from
the correlation between the far-IR and radio luminosity observed in nearby
star-forming galaxies \citep{Condon1992}.

The far-IR luminosity is found by integrating over the fitted modified
black body. The mass $M_\mathrm{dust}$ of dust at $T_\mathrm{dust}$ is
related to far-IR luminosity by 
\begin{equation}
  \label{eq:Mdust}
  L_\mathrm{FIR} = 4\pi\ M_\mathrm{dust}\ \int \kappa(\nu)B_\nu(T_\mathrm{dust})\ d\nu, 
\end{equation}
or to the flux density $S_{\nu_\mathrm{obs}}$ observed at frequency
$\nu_\mathrm{obs}$ (See Eq.~\ref{eq:mbb})

A major source of uncertainty in estimating the dust mass comes from
uncertainties of the mass absorption coefficient $\kappa$, which is
poorly constrained by observations or laboratory experiments.  We
adopt a value of $0.4\ \mathrm{cm^2/g}$ at 1200~$\mathrm{\mu m}$,
which is in the range of values found in the literature \citep[~and
references therein]{Alton2004}. For each source, the derived
temperature, spectral index, luminosity, and dust mass are listed in
Table~\ref{tab:derived}.

The dust spectral index could be fitted for only two of our sources,
APM~08279+5255 and PSS~2322+1944, resulting in $\beta=1.9 \pm 0.3$ and $1.5
\pm 0.3$, respectively. The fixed value of $\beta=1.6$ for the other sources
is based on the mean SED discussed in \S~4.2. Our values are consistent with
those found for dusty local galaxies: $1.3 \le \beta \le 2.0$ \citep[~and
references therein]{Dunne2000, Priddey2001, Alton2004}, and are slightly
lower than the $\beta = 2$ expected for pure silicate and/or graphite grains
\citep{Draine1984}.

\begin{figure}
\plottwo{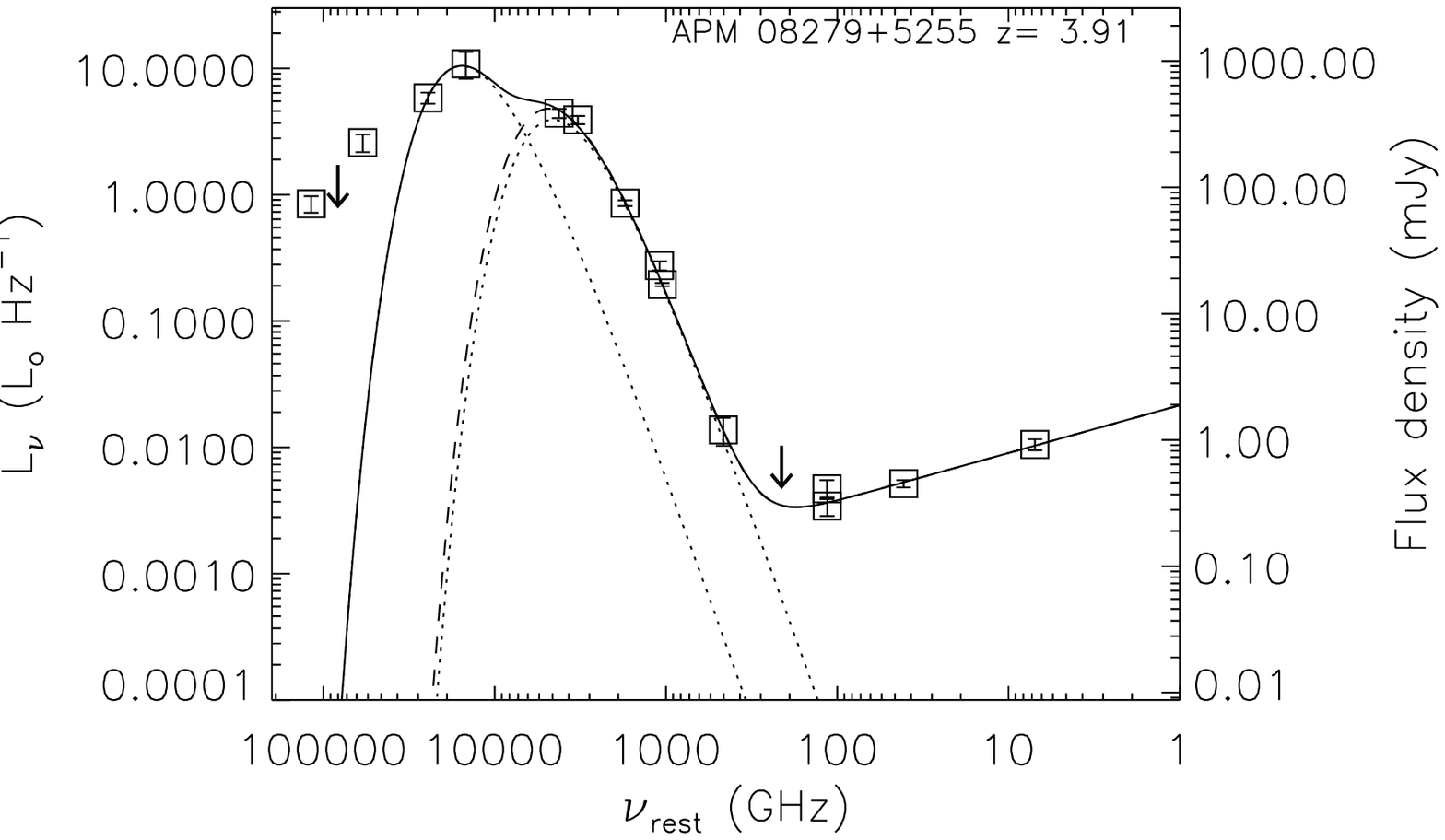}{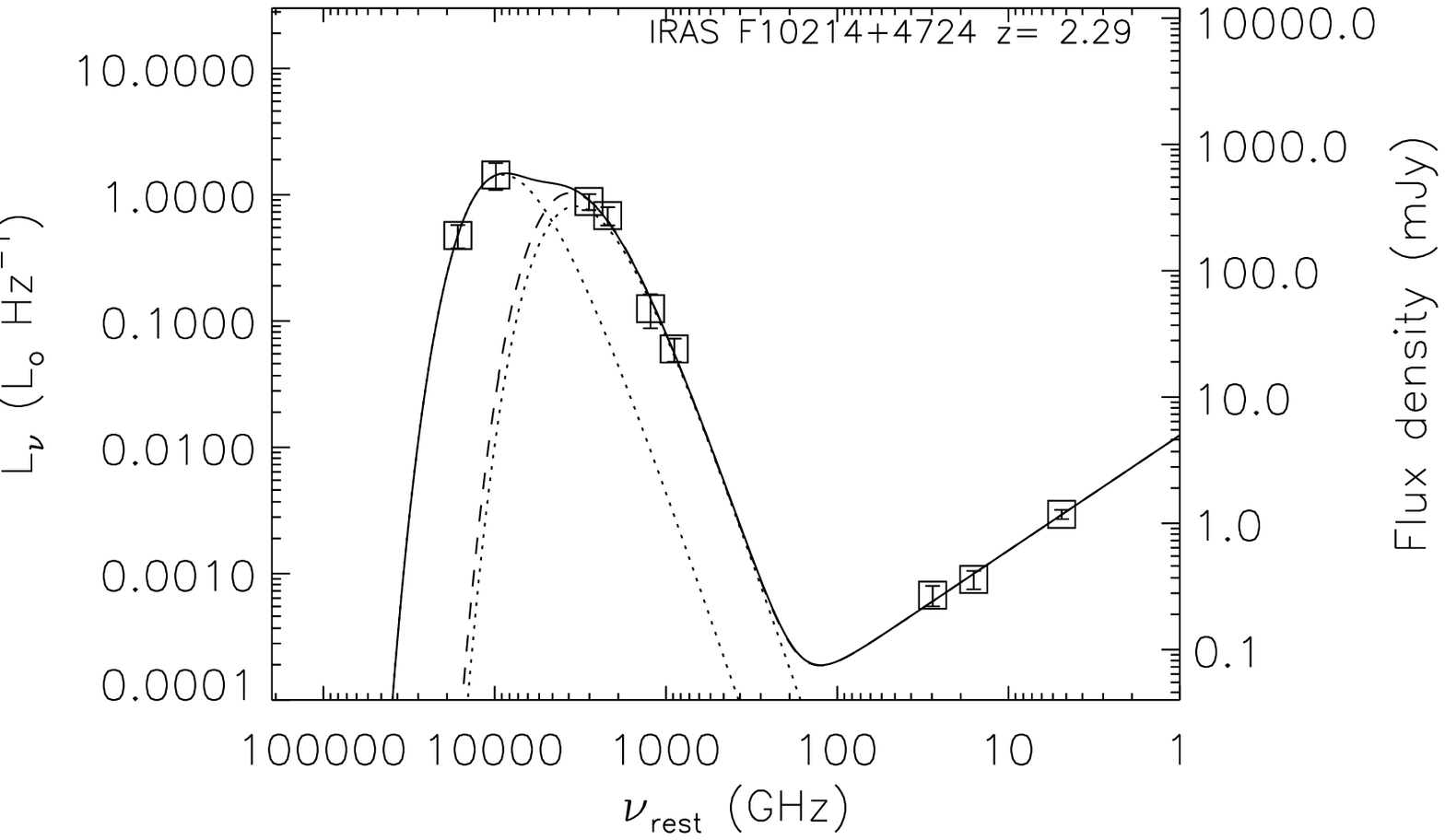} \\
\epsscale{0.5}
\plotone{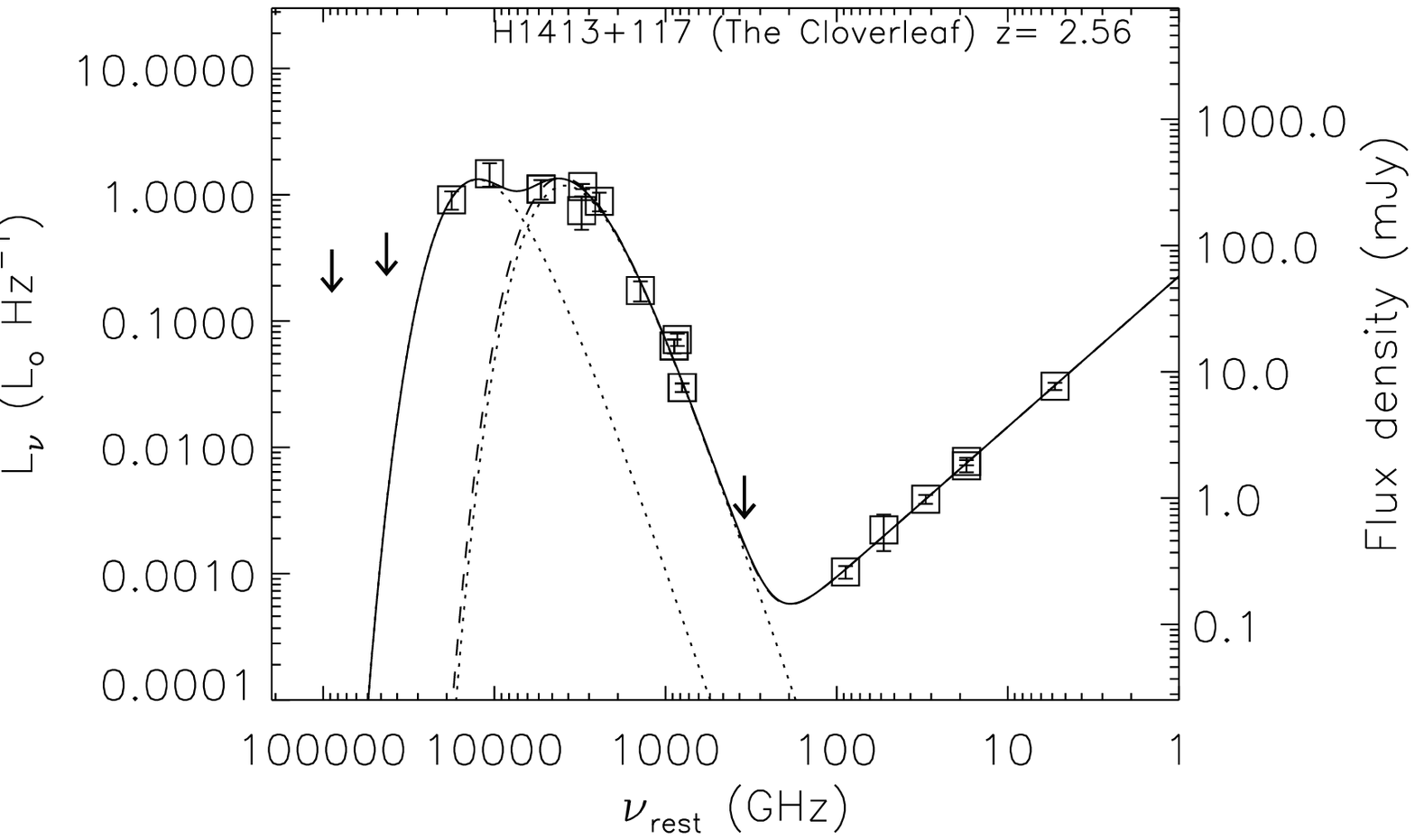}
\caption{SEDs of APM~08279+5255, IRAS~10214+4724 and the Cloverleaf in the
  rest frame of the sources.  The solid line shows the best fit to the
  mid/far-IR data using two components model (plain and dotted line) and to
  the radio using a power-law (see text for parameters). The dashed line
  correspond to a single component fit to the far-IR
  data.\label{fig:two_comp}}
\end{figure}

The derived temperatures for the warm dust are in the range 30 to 60~K,
which is typical for local IR-luminous galaxies, where the heating is
dominated by young massive stars \citep{Dunne2000}.  The values are also
comparable to those found for other high-$z$ sources \citep{Benford1999}.
Our single-component fits do not account for the hot dust component that
dominates at rest frame mid-IR wavelengths.  For APM~08279+5255, e.g., when
adopting $\beta = 1.9$, a two dust component fit to the SED between 10 and
800~\micron\ yields $T_\mathrm{hot} = 155 \pm 17 \ \mathrm{K}$ and
$T_\mathrm{warm} = 45 \pm 2 \, \rm K$. The warm dust temperature well agrees
with the value found from a single dust component fit. The situation is
similar for the Cloverleaf, where a single component fit yields
$T_\mathrm{dust} = 38 \pm 3 \, \rm K$ and $\beta = 2.0 \pm 0.2$, whereas a
two-component fit with fixed $\beta =2$ yields $T_\mathrm{hot} = 118 \pm 13
\ \mathrm{K}$ and $T_\mathrm{warm} = 36 \pm 1 \, \rm K$, which also agrees
with the results of \citet{Weiss2003}. The same applies in the case of
IRAS~F10214+4724 with a single component fit yields to $T_\mathrm{dust} = 44
\pm 7 \, \rm K$ with $\beta$ fixed to 1.6, and a two component fit to
$T_\mathrm{hot} = 93 \pm 12 \ \mathrm{K}$ and $T_\mathrm{warm} = 41 \pm 8 \,
\rm K$. These three examples (see fig~\ref{fig:two_comp}) suggest that the
temperatures found for the warm dust from single component fits are not much
affected by the hot dust component -- although this probably depends
somewhat on the relative intensities of both components. Further
measurements at the rest frame peak emission, from 10 to $100\ \micron$, are
needed to clearly solve this problem, note that a dust temperature
distribution is more likely to be found, instead of a single or two dust
components.

The derived dust masses range from a few $10^8$ to $10^9 \ \mathrm{M_\odot}$
(Table~\ref{tab:derived}), indicating huge reservoirs of gas in the high-$z$
quasars. However, uncertainties in these mass estimates arise from the
assumed value of the mass absorption coefficient $\kappa$, which is
constrained only within a factor of 4 \citep{Alton2004}. Furthermore, there
may well be a cold dust component which emission would remain hidden below
that of the warm component even if it contained up to three times the mass
of the warm component \citep{Dunne2000, Dunne2001}.

\subsection{Dust temperature and spectral index of the mean far-IR SED 
of high-$z$ quasars}
\label{sec:mean}

The determination of both $\beta$ and $T_\mathrm{dust}$ is currently
possible only for a few objects. Even when enough photometric data
points are available for individual sources, the uncertainties on the
temperature and the spectral index are large and the
$\beta-T_{\mathrm{dust}}$ degeneracy prevents reliable estimates of
the far-IR luminosities.

\begin{figure}
\centering
\includegraphics[angle=90,width=0.75\linewidth]{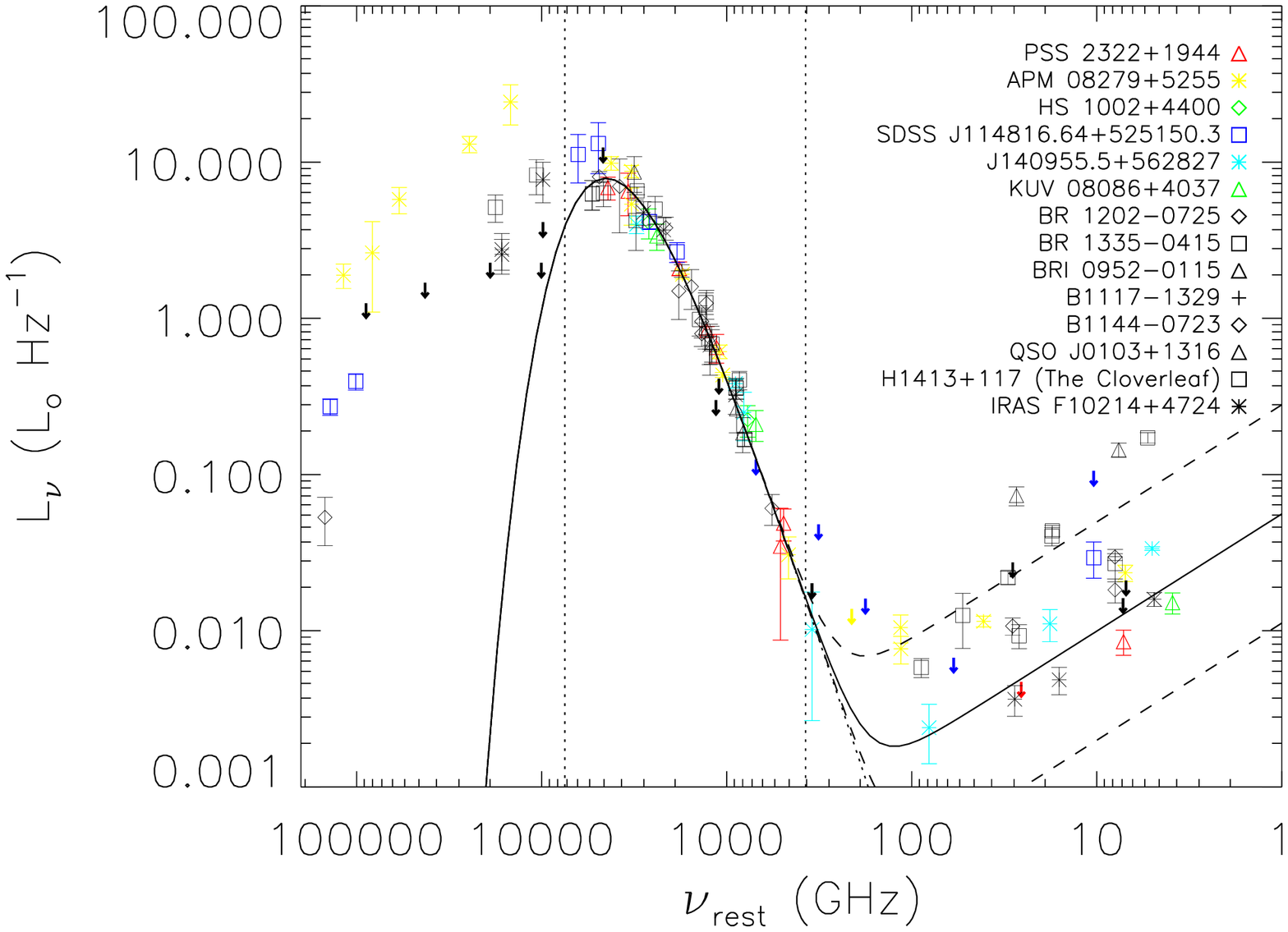}
\caption{Combined SED, in the rest frame, for all the 
  high-$z$ quasars from this paper and sources discussed in
  \citet{Benford1999} and \citet{Priddey2001} (see references therein
  and in Fig.~\ref{fig:indiv_plots}).  The SEDs have been normalized
  to the far-IR luminosity of PSS~2322+1944. Each
  quasar is represented with a different symbol identified in the
  panel. The best fit to the rest frame far-IR data together with the
  derived radio continuum are shown using the same definitions as in
  Fig.~\ref{fig:indiv_plots}.  The corresponding dust temperature and
  spectral index are displayed in Fig.~\ref{fig:combined_fit_chi2}.
  The two vertical dot-dashed lines delineate the wavelength domain
  defined as the far-IR in this paper.
  \label{fig:combined_fit}}
\end{figure}

To better constrain the mean properties of the high-redshift quasar warm
dust component, we have fitted a single temperature grey body to the
photometric points of all quasars with at least two measurements in the
rest frame far-IR.  This provides a useful empirical description of the mean
far-IR SED of dusty high-$z$ galaxies, which can be used as a template to
derive the far-IR luminosities when only a few photometric measurements are
available. The 14 quasars of this sample have redshifts in the range 1.8 to
6.4, which provides a ``wavelength coverage'' of the ensemble SED much
wider than that possible for individual objects.
Figure~\ref{fig:combined_fit} shows the resulting mean SED.

\begin{figure}
\centering
\includegraphics[angle=90,width=0.5\linewidth]{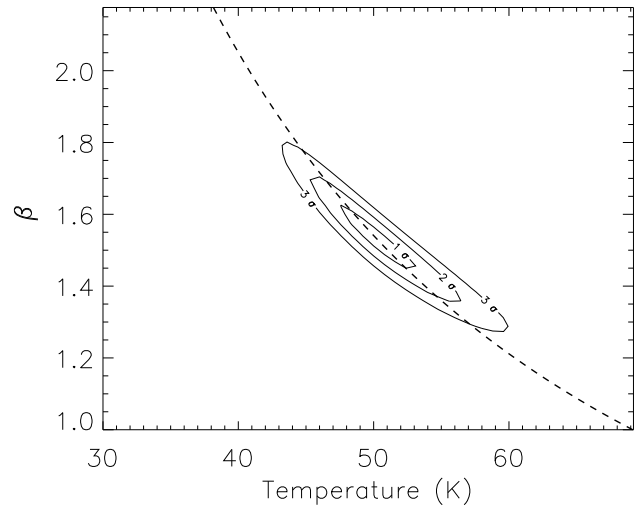}
\caption{Contours of $\chi^2$ in the $\beta - T_\mathrm{dust}$ plane for the
  combined SEDs of the high-$z$ quasars shown in
  Fig.~\ref{fig:combined_fit}. Contours represent the 1, 2, and 3$\sigma$
  uncertainties. The dashed line represents the $\beta - T_\mathrm{dust}$
  degeneracy (see text for details).\label{fig:combined_fit_chi2}}
\end{figure}

A similar analysis was done by \citet{Priddey2001} for a smaller sample of
$z>4$ quasars, where they found $T_\mathrm{dust} = 41 \pm 5 \ \mathrm{K}$
and $\beta = 1.95 \pm 0.3$.  Instead of iteratively normalizing the SEDs, we
left the normalization of each SED a free parameter in the combined fit.
Thereby no prior assumption is made on the scale, but each source SED needs
two or more far-IR flux measurements to bring at least one degree of freedom
to the fit.  The best fit values, $T_\mathrm{dust} = 47 \pm 3 \ \mathrm{K}$,
$\beta = 1.6 \pm 0.1$, the $\chi^2$ contours and the $\beta-T_\mathrm{
  dust}$ degeneracy are shown in Fig.~\ref{fig:combined_fit_chi2}.  The
fitted SED is overplotted on the ensemble of scaled measurements in
Fig.~\ref{fig:combined_fit}.  Although consistent (within $1 \sigma$) with
the results of \citet{Priddey2001}, our findings suggest a somewhat higher
dust mean temperature and lower $\beta$.

The derived mean spectral index of $\beta = 1.6 \pm 0.1$ is not consistent
with the value of $\beta = 2$ expected for dust grains made of pure
silicates and/or graphites \citep{Draine1984}.  However, a range of dust
temperatures will lower the effective measured $\beta$. For example, the sum
of two modified black-bodies with dust temperatures $T_\mathrm{warm}$,
$T_\mathrm{cold}$ and $\beta =2$ provides a fit to the mean far-IR SED that
is as good as the one with a single dust temperature and $\beta=1.5$ - see
also \citet{Dunne2001}.  The mean value of $\beta =1.6$ was adopted in
\S~4.1 to fit the dust temperature for sources with only a few photometric
measurements.  The current data does not well constrain an additional cold
component, for which measurements at longer wavelengths, 2 to 3~mm, would be
most valuable.
 
\subsection{Infrared to Radio spectral index}

For local star-forming galaxies a tight linear correlation is found between
the radio continuum (monochromatic) luminosity, $L_\mathrm{1.4\, GHz}$, and
the far-IR luminosity, $L_\mathrm{FIR}$, with a scatter of a factor $\approx
2-3$ over more than four orders of magnitude in luminosity
\citep{Condon1992, Yun2001}. This tight relation is generally understood to
be due to star-formation activity, measuring the dust heated by young stars,
and the radio synchrotron emission from supernova remnants.  That this
relation also holds at higher redshifts was shown by \citet{Appleton2004},
who report evidence for its validity out to $z=2$ by matching {\em Spitzer
  Space Telescope} 24 and 70~\micron~ sources and VLA radio sources.  In the
following, we check whether the Condon correlation also holds for high-$z$
far-IR-luminous quasars.

The far-IR/radio relationship is usually quantified by a parameter
\begin{equation}
  \label{eq:q_parameter}
  q \equiv \log({L_\mathrm{FIR} \over 3.75 \times 10^{12} \,
  \mathrm{W}}) - \log({L_\mathrm{1.4\, GHz} \over \mathrm{W \,
  Hz^{-1}}}),
\end{equation}
where $L_\mathrm{1.4\, GHz}$ is the monochromatic rest frame 1.4~GHz
luminosity \citep{Helou1985, Condon1992}, $L_\mathrm{FIR}$ is the
far-IR luminosity, which was originally computed from the rest frame
60 and 100~\micron~ IRAS fluxes \citep{Helou1985} under the assumption
of a typical dust temperature of $\approx 30 \, \mathrm{K}$. However, these
definitions are not practical for ULIRGs that have higher dust
temperatures, and we therefore use the integrated far-IR luminosity of
the warm dust component.

To compute the radio luminosities we fitted a power-law to the observed
radio flux densities and extrapolate to the rest frame 1.4 GHz flux
density.  When only one data point is available, we use a radio
spectral index of $-0.75$ \citep{Condon1992}.

\begin{figure*}
\centering
\includegraphics[angle=90,width=0.7\linewidth]{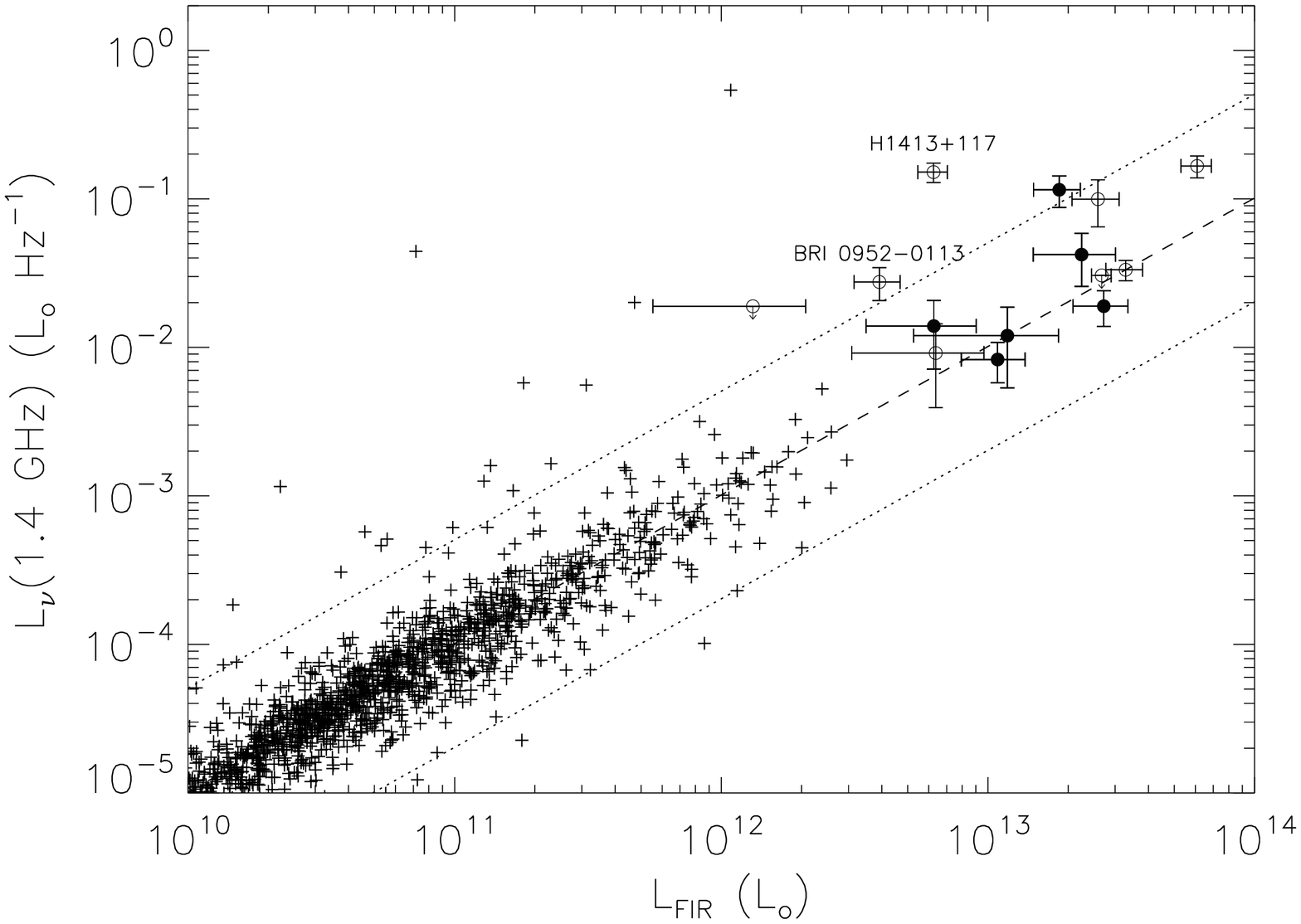}
\caption{rest frame  1.4~GHz luminosity as a function of the 
  $L_\mathrm{FIR}$ as defined by \citet{Condon1992} - see text. The
  crosses are for the IRAS 2 Jy sample of \citet{Yun2001} and the
  circles for the sources discussed in this paper which have radio
  detection (filled circles are sources from
  Table~\ref{tab:observation}, whereas open circles are sources taken
  from the literature and listed in Fig.~\ref{fig:combined_fit}). The
  dashed line shows the mean value of $q$ while the dotted lines
  display the IR and radio excesses which are 5 times above and
  below the value expected from the linear far-IR/radio relation. When
  known, the luminosities of the high-$z$ sources have been corrected
  for lensing.
\label{fig:FIR-radio}}
\end{figure*}

The resulting $q$ values are listed in Table~\ref{tab:derived}.
Figure~\ref{fig:FIR-radio} plots the radio luminosity against the far-IR
luminosity for the 2~Jy galaxy sample of \citet{Yun2001}. To be consistent
with our definitions, we recomputed the far-IR luminosities by fitting a
single dust temperature modified black-body with $\beta=1.6$.  The figure
compares the Yun sample to the high-$z$ quasars sample, except for
HS~1002+4400 where no radio data are available (see the caption of
Fig~\ref{fig:FIR-radio} for details).

Most of our quasars well follow the far-IR/radio correlation well, showing a
median $q = 2.2\pm0.4$, in good agreement with the value of $2.3\pm0.1$
found for the Yun sample. Two quasars show unusually high radio to far-IR
flux ratios: BRI~$0952-0115$ which shows a radio luminosity similar to that
of a low-luminosity radio galaxy \citep{Yun2000} and the Cloverleaf, which
contains a known weakly radio-loud AGN.

The fact that the high-$z$ quasars reasonably follow the Condon
relation for star-forming galaxies suggests that their radio and far-IR
emission do also arise from star formation, although a weak
contribution from the AGN is not excluded. Further evidence for this
is provided by the vast amounts of dense molecular gas detected in CO
line emission for some of the quasars (see \S~\ref{sec:results}) , and
by the implied several kpc spatial extent of these gas reservoirs,
which is very similar to that of the starbursts rings resolved in
nearby Seyfert and starburst galaxies \citep{Garcia-Burillo2003}.

\section{Conclusions}

This paper reports sensitive measurements of 350~\micron\ dust
emission from high-redshift quasars.  The detection of six $1.8 \le z
\le 6.4$ quasars doubles the number of quasars at high redshift for
which 350~\micron\ photometry is available. Combined with observations
at mm/submm wavelengths, the 350~\micron\ data allow
us to sample the rest frame far-IR SEDs of these high-$z$
sources and, thereby, to constrain the properties of their warm dust
emission.  The 350~\micron\ measurements are key in deriving the
far-IR luminosities since they sample the peak of the warm dust
emission.  For the high-$z$ quasars, the derived temperatures are in
the range 40 to 60~K and the luminosities are a few $10^{13} \,
\mathrm{L_\odot}$. The derived dust masses range from a few $10^8$ to
$10^9 \, \mathrm{M_\odot}$.

The mean far-IR SED of all high-$z$ quasars measured with two or more rest
frame far-IR data points is best fit with a grey-body of temperature,
$T_\mathrm{dust} = 47 \pm 3 \, \mathrm{K}$ and a dust emissivity spectral
index, $\beta = 1.6 \pm 0.1$.  However, note that there are considerable
variations and uncertainties in the determination of the dust temperature
and spectral index in individual objects, which is not reflected in the
quoted uncertainties here. To determine accurately $\beta$ and $T_{dust}$, a
good sampling of the SED, especially along the Rayleigh-Jeans part of the
modified blackbody is needed with measurements around the emission peak.
Photometric measurements at high-frequency (e.g., at 350~\micron) are
important to sample the peak of the redshifted far-IR emission and derive
the far-IR luminosity, whereas data at low frequencies (2 or 3~mm) would be
key to constrain $\beta$. Future facilities such as the Atacama Large
Millimeter Array (ALMA) will improve the sensitivity in the submm/mm regime
by more than an order of magnitude relative to current instrumentation and
will enable the measurement of the SEDs of high-$z$ sources over a wide
range in frequency with a precise calibration, including much less luminous
and extreme objects than the ones which can be observed today.

All the radio-quiet high-$z$ quasars studied in this paper approximately
follow the far-IR/radio correlation, showing a median $q = 2.2 \pm 0.4$,
consistent with that found for local star-forming galaxies. This result is a
further indication that, in these high-$z$ radio-quiet quasars, the radio
and far-IR emission does arise from star formation.

The warm (40-60~K) dust component found from single component fits to the
SEDs of high redshift quasars is not much affected by the presence of a hot
(several 100~K) dust component that dominates at rest frame mid-IR
wavelengths, as shown by the examples of APM~08279+5255 and the Cloverleaf.
The warm dust component is thus found to be a relatively good tracer of the
starburst activity of the quasars's host galaxy.  More mid-IR measurements
are needed to assess the degree to which the warm and hot dust components
can be treated independently in high-$z$ quasars.  In this respect, the {\em
  Spitzer Space Telescope} will provide valuable photometric data at mid-IR
wavelengths and allow us to build complete SEDs for a number of high
redshift quasars from their rest frame near- to far-IR wavelengths to
investigate the relative importance of the AGN and starburst activity.



\acknowledgments

The CSO is funded by the NSF under contract AST02-29008. A.B.  acknowledges
financial support from the Programme National de Galaxies. We thank A. Jones
and F. Boulanger for fruitful discussions on the subject of this work.

\clearpage

\clearpage

\clearpage

\clearpage

\clearpage

\clearpage

\input{tab1} \input{tab2}

\end{document}

%% file: tab1.tex
\begin{table}
\scriptsize
\begin{center}
\caption{Observational Parameters.\label{tab:observation}}
\begin{tabular}{lccllccc}
\tableline\tableline
\multicolumn{1}{c}{Source} & 
\multicolumn{1}{c}{$z$} &
\multicolumn{1}{c}{$M_\mathrm{B}$} &
\multicolumn{1}{c}{R.A.} &
\multicolumn{1}{c}{Dec.} &
\multicolumn{1}{c}{$S_\mathrm{1.2mm}$} &
\multicolumn{1}{c}{$S_\mathrm{350\mu m}$} &
\multicolumn{1}{c}{Int. time} 
\\
  & & &
\multicolumn{2}{c}{(J2000.0)} &
\multicolumn{1}{c}{(mJy, $\pm1 \sigma$)} &
\multicolumn{1}{c}{(mJy, $\pm1 \sigma$\tablenotemark{\dagger})} &
\multicolumn{1}{c}{(min)}
\\
\tableline
KUV~08086+4037                         &  1.78 & $-27.0$   & 08 12 00.41 & 40 28 15.00 & $~4.3 \pm 0.8$\tablenotemark{a} & $ ~69\pm  11$                  &  180  \\
APM~08279+5255\tablenotemark{\ddagger} &  3.91 & $-32.9$   & 08 31 41.70 & 52 45 17.35 & $~~24 \pm ~~2$\tablenotemark{b} & $ 386\pm  32$                  &   20  \\
HS~1002+4400                           &  2.08 & $-28.3$   & 10 05 17.45 & 43 46 09.30 & $~4.2 \pm 0.8$\tablenotemark{a} & $ ~77\pm  14$                  &  170 \\
J1148+5251                             &  6.42 & $-28.4$   & 11 48 16.64 & 52 51 50.30 & $~5.0 \pm 0.6$\tablenotemark{c} & $ ~21\pm  ~6$\tablenotemark{+} &  430 \\
J1409+5628                             &  2.58 & $-28.4$   & 14 09 55.56 & 56 28 26.50 & $10.7 \pm 0.6$\tablenotemark{a} & $ 112\pm  12$                  &  220 \\
PSS~2322+1944                          &  4.12 & $-28.1$   & 23 22 07.25 & 19 44 22.08 & $~9.6 \pm 0.5$\tablenotemark{d} & $ ~79\pm  11$                  &  150 \\
\tableline
\end{tabular}
\tablenotetext{\dagger}{The absolute calibration uncertainty of 20\% is not
  included in the quoted values}
\tablenotetext{\ddagger}{For APM~08279+5255, the flux density measured at
  450~\micron~ with {\sc Sharc II} is $\rm 342 \pm 26 \, mJy$}
\tablenotetext{+}{The flux density for J1148+5251 is derived with a FWHM of
  $11\farcs5$.}
\tablenotetext{a}{\citet{Omont2003}}
\tablenotetext{b}{\citet{Irwin1998} at 1.35~mm}
\tablenotetext{c}{\citet{Bertoldi2003a}}
\tablenotetext{d}{\citet{Omont2001}}
 \end{center}
\end{table}


%% file: tab2.tex
\begin{table}
\small
\begin{center}
\caption{Derived Properties.\label{tab:derived}}
\begin{tabular}{lcccccc}
\tableline\tableline
\multicolumn{1}{c}{Source} & 
\multicolumn{1}{c}{T$\rm _{dust}$} &
\multicolumn{1}{c}{$\beta$} & 
\multicolumn{1}{c}{amp. factor} &
\multicolumn{1}{c}{$L_\mathrm{FIR}$} &
\multicolumn{1}{c}{Dust Mass} & 
\multicolumn{1}{c}{$q$} 
\\
  & 
\multicolumn{1}{c}{(K)} & &  &
\multicolumn{1}{c}{($10^{13}\ \mathrm{L_{\odot}}$)} & 
\multicolumn{1}{c}{($10^{8}\ \mathrm{M_{\odot}}$)} &

\\

\tableline
KUV~08086+4037 & $ 32 \pm   5$ & $1.6^+$      &  --               &  $0.6\pm0.3$  & 23.8 & $ 2.1 \pm  0.2$  \\
APM~08279+5255 & $ 47 \pm   7$ & $1.9\pm0.3$  & $7^\mathrm{[1]}$   & $2.7\pm0.7$  &  5.7 & $ 2.6 \pm  0.1{\dagger}$  \\
  HS~1002+4400 & $ 38 \pm   7$ & $1.6^+$      &  --               &  $1.2\pm0.7$  & 17.3 &                   \\
    J1148+5251 & $ 55 \pm   5$ & $1.6^+$      &  --               &  $2.2\pm0.7$  &  4.5 & $ 2.2 \pm  0.2$  \\
    J1409+5628 & $ 35 \pm   2$ & $1.6^+$      &  --               &  $1.8\pm0.4$  & 48.8 & $ 1.4\pm  0.2^{\ddagger}$ \\
 PSS~2322+1944 & $ 47 \pm   8$ & $1.5\pm0.3$  & $3.5^\mathrm{[2]}$ & $1.1\pm0.3$  &  6.8 & $ 2.5 \pm  0.1$  \\
\tableline
\end{tabular}

\tablenotetext{}{The far-IR luminosities and dust masses are corrected for
  lensing amplification when indicated. \\  The references to the amplification
  factor are : [1] \citet{Lewis2002}, [2] \citet{Carilli2003}. }
\tablenotetext{}{$^{+}$ fixed value} \tablenotetext{}{$^{\dagger}$
  $\alpha_\mathrm{radio} = -0.35\pm0.07$} \tablenotetext{}{$^{\ddagger}$
  $\alpha_\mathrm{radio} = -1.0\pm 0.1$ }
 \end{center}
\end{table}

%% file: ms.bbl
\begin{thebibliography}{}
\bibitem[Alton et al.(2004)]{Alton2004} Alton, P.~B., Xilouris, E.~M.,
  Misiriotis, A., Dasyra, K.~M., \& Dumke, M.\ 2004, \aap, 425, 109
\bibitem[Appleton et al. (2004)]{Appleton2004} Appleton, P.N., Fadda,
  D.T., Marleau, F.R. et al. 2004, \apjs, 154, 147 
\bibitem[Barvainis \& Ivison (2002)]{Barvainis2002} Barvainis, R.~\&
  Ivison, R.\ 2002, \apj, 571, 712
\bibitem[Beelen et al.(2004)]{Beelen2004} Beelen, A., Cox, P., Pety,
  J. et al.  2004, \aap, 423, 441
\bibitem[Benford et al.(1999)]{Benford1999} Benford, D.J., Cox, P.,
  Omont, A., Phillips, T.G., \& R.G. McMahon 1999, \apj, 518, L65
\bibitem[Bertoldi et al.(2003a)]{Bertoldi2003a} Bertoldi, F., Carilli,
  C.L., Cox, P.  et al. 2003a, \aap, 406, L55
\bibitem[Bertoldi et al.(2003b)]{Bertoldi2003b} Bertoldi, F., Cox, P.,
  Neri, R., et al.  2003b, \aap, 409, L47
\bibitem[Blain, Barnard, \& Chapman (2003)]{Blain2003} Blain, A.W.,
  Barnard, V.E., \& Chapman, S.C. 2003, \mnras, 338, 773
\bibitem[Carilli et al.(2001)]{Carilli2001} Carilli, C.~L., et al.\
  2001, \apj, 555, 625
\bibitem[Carilli et al.(2002)]{Carilli2002} Carilli, C.L., Cox, P,
  Bertoldi, F. et al.  2002, \apj, 575, 145
\bibitem[Carilli et al.(2003)]{Carilli2003} Carilli, C.L., Lewis,
  G.F., Djorgovski, S.G.  et al. 2003, Science, 300, 773
\bibitem[Carilli et al.(2004)]{Carilli2004} Carilli, C.~L., et al.\ 
2004, \aj, 128, 997 
\bibitem[Carilli et al.(2005)]{Carilli2005} Carilli, C.~L., et al.\ 
2005, \apj, 618, 586 
\bibitem[Charmandaris et al.(2004)]{Charmandaris2004} Charmandaris, V., 
et al.\ 2004, \apjs, 154, 142 
\bibitem[Condon(1992)]{Condon1992} Condon, J.~J.\ 1992, \araa, 30, 575
\bibitem[Cox et al.(2002)]{Cox2002} Cox, P., Omont, A., Djorgovski,
  S.G. et al.  2002, \aap, 387, 406
\bibitem[Darling \& Wegner(1996)]{Darling1996} Darling, G.~W.~\&
  Wegner, G.\ 1996, \aj, 111, 865
\bibitem[Dowell et al.(2003)]{Dowell2003} Dowell, C.~D., Allen, C.A.,
  Babu, R. et al.  2003, \procspie, 4855, 73
\bibitem[Downes et al.(1999)]{Downes1999} Downes, D., Neri, R.,
  Wiklind, T., Wilner, D.~J., \& Shaver, P.~A.\ 1999, \apjl, 513, L1
\bibitem[Dunne et al.(2000)]{Dunne2000} Dunne, L., Eales, S., Edmunds,
  M., Ivison, R., Alexander, P., \& Clements, D.~L.\ 2000, \mnras,
  315, 115
\bibitem[Dunne \& Eales(2001)]{Dunne2001} Dunne, L.~\& Eales, S.~A.\
  2001, \mnras, 327, 697
\bibitem[Draine \& Lee (1984)]{Draine1984} Draine, B.~T.~\& Lee,
  H.~M.\ 1984, \apj, 285, 89
\bibitem[Fan et al.(2003)]{Fan2003} Fan, X., et al.\ 2003, \aj, 125,
1649
\bibitem[Garc{\'{\i}}a-Burillo et al.(2003)]{Garcia-Burillo2003} 
  Garc{\'{\i}}a-Burillo, S., Combes, F., Hunt, L.K. et al.\ 2003, 
  \aap, 407, 485 
\bibitem[Greve et al.(2005)]{Greve2005} Greve, T.~R., Bertoldi, F., Smail,
  Ian, et al. \ 2005, \mnras, 359, 1165
\bibitem[Hagen et al.(1999)]{Hagen1999} Hagen, H.-J., Engels, D., \&
  Reimers, D. 1999, \aaps, 134, 483
\bibitem[Helou, Soifer, \& Rowan-Robinson(1985)]{Helou1985} Helou, G.,
   Soifer, B.~T., \& Rowan-Robinson, M.\ 1985, \apjl, 298, L7
\bibitem[Hughes et al.(2002)]{Hughes2002} Hughes, D.~H., Aretxaga, I.,
  Chapin, E. L. et al.\  2002, \mnras, 335, 871 
\bibitem[Irwin et al.(1998)]{Irwin1998} Irwin, M.~J., Ibata, R.~A.,
  Lewis, G.~F., \& Totten, E.~J.\ 1998, \apj, 505, 529
\bibitem[Leong et al. (2005)]{Leong2005} Leong M.\ M.\ 2005, URSI Conf. Sec. 
J3-10, 426, \url{http://astro.uchicago.edu/ursi-comm-J/ursi2005/rf-telescope-fabrication/leong}
\bibitem[Lewis et al.(1998)]{Lewis1998} Lewis, G.~F., Chapman, S.~C.,
  Ibata, R.~A., Irwin, M.~J., \& Totten, E.~J.\ 1998, \apjl, 505, L1
\bibitem[Lewis et al.(2002)]{Lewis2002} Lewis, G.~F., Carilli, C.,
  Papadopoulos, P., \& Ivison, R.~J.\ 2002, \mnras, 330, L15
\bibitem[Maiolino et al.(2005)]{Maiolino2005} Maiolino, R., et al.\ 2005,
  \aap, 440, L51
\bibitem[Omont et al.(2001)]{Omont2001} Omont, A., Cox, P., Bertoldi,
  F. et al. 2001, \aap, 374, 371
\bibitem[Omont et al.(2003)]{Omont2003} Omont, A., Beelen, A.,
  Bertoldi, F. et al.  2003, \aap, 398, 657
\bibitem[Petric et al.(2005)]{Petric2005} Petric, A., Carilli, C.L.,
  Bertoldi, F. et al. 2004, \aj, submitted
\bibitem[Priddey \& McMahon (2001)]{Priddey2001} Priddey, R.S., \&
  McMahon, R.G. 2001, \mnras, 324, L17
\bibitem[Robson et al. (2004)]{Robson2004} Robson, I., Priddey, R.S.,
  Isaak, K.  \& McMahon, R.G. 2004, \mnras, 351, L29
\bibitem[Spergel et al.(2003)]{Spergel2003} Spergel, D.~N., et al.\
  2003, \apjs, 148, 175
\bibitem[Soifer et al.(2004)]{Soifer2004} Soifer, B.~T., et al.\ 
2004, \apjs, 154, 151 
\bibitem[Walter et al.(2003)]{Walter2003} Walter, F., Bertoldi, F.,
  Carilli, C.L.  et al.\ 2003, \nat, 424, 406
\bibitem[Wei{\ss} et al. (2003)]{Weiss2003} Wei{\ss}, A., Henkel, C.,
  Downes, D., \& Walter, F.\ 2003, \aap, 409, L41
\bibitem[Wiklind(2003)]{Wiklind2003} Wiklind, T.\ 2003, \apj, 588, 736 
\bibitem[Yun et al.(2000)]{Yun2000} Yun, M.~S., Carilli, C.~L.,
  Kawabe, R., Tutui, Y., Kohno, K., \& Ohta, K.\ 2000, \apj, 528, 171
\bibitem[Yun, Reddy, \& Condon(2001)]{Yun2001} Yun, M.~S., Reddy,
  N.~A., \& Condon, J.~J.\ 2001, \apj, 554, 803
\end{thebibliography}
